*Long-term sedimentary earthquake records along the* the northern branch of the North Anatolian Fault in the Sea of Marmara (NW Türkiye)

**M. Namık Çağatay (1) , Demet Biltekin (2) , Nurettin Yakupoğlu (1) , Emin Güngör , Nurdan Güngör , Gülsen Uçarkuş (1) , Pierre Henry (3) , Alina Polonia (4) , Luca Gasperini (5) , Celine Grall (6, 7) , Dursun Acar , Umut Barış Ülgen , Christos Tsabaris , Asen Sabuncu**

**1 ITÜ - Istanbul Technical University**
**2 Maden Fakültesi = Faculty of mines [Istanbul]**
**3 CEREGE - Centre Européen de Recherche et d'Enseignement des Géosciences de l'Environnement**
**4 ISMAR - Istituto di Science Marine**
**5 ISMAR - Istituto di Scienze Marine [Bologna]**
**6 LIENSs - LIttoral ENvironnement et Sociétés**
**7 LDEO - Lamont-Doherty Earth Observatory**

**This is the accepted version and may slightly differ from the published version.**

**Abstract**

Long-term (geological) earthquake records are important for probabilistic earthquake risk assessment. Such records can be obtained from the study of turbidites triggered by seismic activity in marine and lake basins. The Sea of Marmara (SoM), located on the North Anatolian Fault (NAF), is an important laboratory for the study of paleoearthquake records. This is because it has 2500 years of historical earthquake records with which radiometrically dated sedimentary earthquake records can be correlated, and its deep subbasins have high sedimentation rates (~3 mm /yr) so that individual events can be distinguished. After the destructive 1912 Mw 7.4 Şarköy-Mürefte and 1999 Mw 7.4 Izmit and Mw 7.2 Düzce earthquakes, the main submerged part of the NAF in the SoM represents a seismic gap, where the long-term earthquake history is of cruscial importance for earthquake risk assessment.

We have studied nine cores recovered along the most active northern strand of the NAF (i.e. the

Main Marmara Fault: MMF), using high resolution digital X-Ray Radiography and μ-XRF Core Scanner, MSCL physical properties and grain-size analyses. The chronology was determined using accelerated mass spectroscopy (AMS) radiocarbon and radionuclide methods. In the cores, coseismic turbidites commonly consists of a basal part with multiple sand-silt laminae above a sharp and often erosional base and an overlaying graded mud part (homogenite). The basal parts have high gamma density and magnetic susceptibility, and are often enriched in one or more of elements, such as Si, K, Fe, Ti, Zr, Ca and Sr, indicative of coarse siliciclastic and carbonate shell debris.

Radionuclide and radiocarbon dated coseismic turbidites in different SoM subbasins extending back to over more than 5000 yrs provide average earthquake recurrence time between 220 and 300 yrs along the different segments of the MMF. These results are compatible with GPS velocities and geological slip rates. However, the intervals between two consecutive events vary widely between 50 to 695 years for the different northern NAF segments.

Integration of our results with the previous studies indicate that the last M>7 earthquake events recorded are the 1999, 1509, 1296 and 740 earthquakes on the İzmit Gulf segment; the 1894, 1509, 1343, 1090, 740 earthquakes on Prince Islands segment in the Çınarcık Basin ; 1894, 1766, 1509, 989, 869 or 862, 740, 447 earthquakes on the Central High segment and in the Central Basin; and the 1912, 1766, 1344, 989, 447 earthquakes on the western Marmara segment. The 1912 Şarköy-Mürefte earthquake record is absent in the Central Basin, which suggests that the earthquake rupture did not extend beyond east of the Central High. The distribution of the last events indicates that the most earthquake-prone part of MMF is the Central High segment located SW of Istanbul.

**Keywords**: Sea of Marmara, earthquake records, earthquake risk, recurrence time, radiocarbon dating, radionuclides

## 1. Introduction

Long-term paleoseismological information on active faults such as earthquake recurrence interval and timing of the last event are important for the probabilistic earthquake risk assessment. In recent years, subaquaous paleoseismological investigations using sediment cores as archives of long-term earthquake records have become a common practice in seismically active marine basins of different tectonic settings. These include the subduction related settings such as the Ionian Sea, Cascadia, Japon Trench and Hikurangi margins (e.g., Pouderoux et al., 2012; Goldfinger et al., 2009, 2012, 2015; Patton et al., 2013; Gutiérrez-Pastor et al., 2013; Nelson et al., 2014; Polonia et al., 2015; Nakamura etal., 2013; Strasser et al., 2013; Ikehara et al., 2016; Kioka et al., 2019a; McHugh et al., 2016; Usami et al., 2018; Molenaar et al., 2019), as well as transtensional basins developed along transform plate boundaries such as the Sea of Marmara (SoM) (e.g., Sarı and Çağatay, 2006; McHugh et al., 2006; 2014; Drab et al., 2012; 2015a; Çağatay et al., 2012, Beck et al., 2015; Yakupoğlu et al., 2019, 2022).

The SoM is ideal for paleoseismological studies for three main reasons: i) its location on a major continental transform boundary, the North Anatolian Fault (NAF), that has created Mw ≥7 historical earthquakes (Fig. 1A), ii) high sedimentation rates (SR) (1-3 m/ka) in up to 1270 m-deep transtenional subbasins so that individual events can be separated, and iii) nearly 2500 year-long historical earthquake records with which carefully dated sedimentary records could be matched.

In the Sea SoM region, the NAF splits into branches in a ~100 km wide deformation zone, mainly because of the N-S extensional regime of the Aegean (Okay et al., 2000; İmren et al., 2001; Armijo et al., 1999, 2002, 2005; Şengör et al., 2005; 2014; Le Pichon et al., 2001, 2015; Çağatay and Uçarkuş, 2019) (Fig. 1A). The most active northern branch, the Main Marmara Fault (MMF, Le Pichon et al., 2001), cuts the SoM floor in an east-west direction (Fig. 1B-G), and accommodates ~75 % of total 24 mm/yr dextral strike-slip motion along the Eurasian-Anatolian plate boundary (Straub and Kahle, 1997; Hubert-Ferrari et al., 2002; McClusky et al., 2000; Meade et al., 2002; Flerit et al., 2003; Reilinger et al., 2006). The NAF has created devasting Mw>7 earthquakes with a repeat interval of ~250 years according to the historical records, which show the occurrence of 55 large (Ms>6.8) earthquakes and 30 tsunamis in the Marmara region in the last 2000 years (Guidoboni, 1994; Ambraseys and Finkel, 1995; Ambraseys, 2002; Yalçıner et al., 2002; Guidoboni and Comastri, 2005; Altınok et al., 2011).

The submerged MMF consists of several segments (Armijo et al., 2002, 2005; Oglesby et al., 2008; Çağatay and Uçarkuş, 2019). From east to west, these are the İzmit Gulf, Prince Islands, Central High (Kumburgaz), Western Marmara and Ganos segments (Fig. 1). With the İzmit Gulf and Ganos segments ruptured during the 1999 Mw7.4 İzmit and 1912 Mw7.4 Şarköy-Mürefte earthquakes respectively, a major part of the submerged MMF constitutes a seismic gap which is expected to rupture in the next few decades producing one or more earthquakes with Mw ≥7 (Hubert-Ferrari et al., 2002; Parsons et al., 2000; Le Pichon et al., 2001; Parsons, 2004; Oglesby et al., 2008; Murru et al., 2016).

Mass-transport deposits including turbidites constitute a significant part of the 5-6 km-thick sedimentary sequence in the ~1250 m deep SoM basins, most of which are considered to have been seismically triggered, and as such, correlated with the historical earthquakes (Sarı and Çağatay et al., 2006; McHugh et al., 2006, 2014; Beck et al., 2007, 2015; Eriş et al., 2012; Çağatay et al., 2012; Drab et al., 2012, 2015a; Yakupoğlu et al., 2019, 2022). Despite the several previous subaqueous paleoseismological studies in the SoM, there are gaps in terms of geographical coverage to unravel the long-term seismic behaviour and to assess the seismic risk of the different MMF segments. In addition, further studies are needed on the sedimentology, physical and geochemical properties of coseismic turbidites to gain insights into their transport and depositional mechanisms in the SoM.

In this paper, we attempt to address these issues by integrating our new core data with previously published paleoseismological and historical data. Our new data come from the analyses of four piston cores and five water/sediment interface cores, located along the MMF in the Çınarcık Basin, Central High, Central Basin, Western High and Tekirdağ Basin. The cores were visually described and analysed for physical properties, XRF elemental geochemistry and grainsize distribution, and age dated by AMS radiocarbon and radionuclide techniques. For a complete synthesis and discussion of the seismic risk assessment, we integrated our data with the published paleo-earthquake data by Mchugh et al. (2006, 2014), Çağatay et al (2012), Drab et al. (2012, 2015a) and Yakupuğlu et al. (2019, 2022).

## 2 Study area

### 2.1 Morphology, hydrography and oceanography

The SoM is a 280-km long and 80-km wide sea between the Mediterranean Sea and the Black Sea (Fig. 1A). It is connected to the adjacent seas by the Dardanelles (Çanakkale) and Bosporus (İstanbul) straits, with sill depths of 65 and 35 m, respectively. The SoM consists of 90-100 m-deep shelf areas, three ~1250 m-deep basins (Çınarcık, Central and Tekirdağ), 450-600 m-deep NE-trending highs (Central and Western) separating the deep basins in between. The other notable morphological features are the ~825 m-deep Kumburgaz Basin on the Central High, ~400 m-deep İmralı Basin in the south, and E-W oriented 100-200 m-deep İzmit and Gemlik gulfs in the east (Fig. 1A).

The northern shelf is narrower (<20 km) than the southern shelf (up to 45 km). The slopes connecting the shelf areas to the deep basins are steep, with slope angles of up to 29° and 16° in the north and south margins, respectively. The slopes are marked by submarine canyons and landslide scars (Zitter et al., 2012; Çağatay et al., 2012; Çağatay and Uçarkuş, 2019). The major canyons are the İzmit, North İmralı and the Bosporus canyons in Çınarcık Basin, Şarköy Canyon in the Tekirdağ Basin, and Büyükçekmece Canyon in the Kumburgaz Basin.

A two-way water exchange occurs through the SoM and its connecting straits, with Mediterranean waters (salinity: ~38.5 psu) and Black Sea waters (~18 psu), forming the lower and upper currents, respectively (Beşiktepe, et al., 1994; Chiggiato et al., 2012; Aydoğdu et al., 2018). A permanent pycnocline at -25 m prevails between the two water masses. Because of mixing between the lower and upper currents during the passage through the relatively narrow Bosporus and Dardanelles straits, the salinity of the surface and bottom waters are modulated to ~38 psu and ~22 psu (Beşiktepe et al., 1994). The lower and upper water masses in the SoM have residence time of 4-5 months and 6-7 years, respectively. The oxygen content of the bottom waters in the deep Marmara basins decrease from 50 µmol/kg in the Tekirdağ basin to 8 µmol/kg in the Çınarcık Basin (Tryon et al., 2012). The bottom waters are occasionally hypoxic to anoxic in the semi-enclosed İzmit and Gemlik gulfs, because of the slow circulation and heavy municipal and industrial inputs (Tolun et al., 2001; Balkıs, 2003; Yüksek et al., 2002; Giuliani et al., 2017).

The main freshwater and sediment inputs are provided mainly by Kocaçay, Gönen and Biga rivers from the southern drainage basin. These rivers provide the SoM with a total of 5.6 km$^3$/yr freshwater and 2.2x10$^6$ tons/yr suspended sediment (EİE, 1993). Most of this sediment load is deposited over the southern shelf. There is only minor sediment input to the northern shelf from the small northern drainage basin (Okay and Ergün, 2005).

The SoM was a fresh-brackish water lake during the last glacial, before the reconnection and establishment of marine conditions at 12.55 ka BP (Çağatay et al., 2015). The water level during the late glacial was controlled by mainly by the sill depth of the Dardanelles Strait, which was at 85 m just before the reconnection and changed in tandem with global sea level thereafter (Çağatay et al., 2009).

## 2.2  Morphotectonics of the Sea of Marmara along the Main Marmara Fault

In this section, we decsribe from east to west the morphotectonic regions of the Sea of Marmara from which the studied cores were recovered, except for İzmit Gulf where discussion on the paleoseisology is based on data previously published by Çağatay et al. (2012), McHugh et al. (2006) and Bertrand et al. (2011).

### 2.2.1  İzmit Gulf

The İzmit Gulf is 48 km long and 2-10 km wide  E-W elongated inlet in the east of the SoM, (Fig. 1B). It consists of Western (Darıca), Karamürsel and Gölcük basins. The Karamürsel Basin in the centre is the deepest part of the Gulf with a maximum depth of 210 m (Çağatay et al., 2012; Kurt and Yücesoy, 2009). It is connected with the adjacent basins via the Hersek and Gölcük sills having sill depths of 55 and 38 m, respectively. The Karamürsel Basin has a small drainage area with two small streams in the south, which form the Çatalburun and Hersek delta-fan complexes (Fig. 1B). The formations in the Karamürsel basin catchment include the Neogene, Eocene and Paleozoic silciclastic rocks and Eocene andesites in the south, and the Paleozoic siliciclastics and Cretaceous carbonates in the north. The Gulf is dissected by the İzmit Gulf segment of the MMF, which during the Mw 7.4 1999 İzmit earthquake, ruptured along the full length of the gulf from Gölcük to offshore Yalova in the Çınarcık Basin (Barka et al. 2002; Uçarkuş et al., 2011).

### 2.2.2 Çınarcık Basin

The Çınarcık Basin is a ~50 km-long and 12 km wide (average), wedge-shaped depression with a maximum depth of -1270 m and an area of 575 km$^2$ (Fig. 1C) (Çağatay and Uçarkuş, 2019). It is bounded in the north by the transtentional WNW-trending Prince Islands segment, which forms a ~1000 m-high steep (25-29°) escarpment. South of İstanbul, this fault connects with the E-trending Central High segment. The northern escarpment is marked with submarine canyons and a bench at its base. The bench is characterized by transtensional right-lateral fault deformation (Fig. 1C). The basin is bordered by the oblique strike-slip Armutlu Fault in the south and the transpressive Central High in the west. To the east, there are two large submarine landslides near the entrance of the İzmit Gulf and İzmit Canyon. The southern slope of the Çınarcık Basin is less steep (15–20°), and marked by a number of WNW-trending extensional en-echelon faults and the sinuous North İmralı Canyon (Figs. 1C, 2A) (Çağatay et al., 2014). The main source of sediments in the Çınarcık Basin is the south drainage area of the SoM (EİE, 1993), and İmralı Canyon appears to be the most important conduit for sediment transport and turbidity currents. This is supported by similar heavy mineral composition of the sands from the Kocasu River's delta and turbidites in the Çınarcık Basin cores, which are enriched mainly in amphibole and pyroxene followed by subequal amounts (< 15%) of garnet epidote, chlorite and opaques (Okay and Ergün, 2005).

The depocentre of the basin is located in the east of the basin, close to the the active Prince Islands Fault (Fig. 2A). The recent sedimentary succession in the basin gently dip and thicken nothwards, towards the active fault, and thin and wedge out towards the southern margin, where it is intercepted by the en-echelon extensional faults (Figs. 1C, 2A).

### 2.2.3 Central High

The Central High is a 40 km wide NE-trending transpressive structure between the Çınarcık and Central basins (Fig. 1D) (Okay et al., 2000; Şengör et al., 2014; Çağatay and Uçarkuş, 2019). It has an irregular relief rising up to ~ -450 m, about ~800 m above the deep basins. The high hosts the Kumburgaz Basin with a maximum depth of ~850 m, which is bound to the north and south by active faults and by a steep continental slope to the north (Fig. 2B). The basin has a high SR (2-2.5 mm/yr) for Holocene period (Beck et al., 2007; Yakupoğlu et al., 2019), with sediments mostly supplied via the Büyükçekmece Canyon (Fig. 1D). The SR elsewhere on the Central High is low (0.1-0.5 mm/yr) compared to the Kumburgaz Basin.

The Central High is cut by the Central High segment of the NAF. Micro-bathymetric mapping and observations by Victor 6000 ROV and Nautile manned submersible dives show that the fault furrow is covered by sediments without a fresh fault scarp, suggesting a lack of recent activity (Armijo et al., 2005; Henry et al., 2008). The submersible observations also show thermal hydrocarbon gas emissions from the fault, originating from the Thrace Basin (Géli et al., 2008; Bourry et al., 2009; Burnard et al., 2012; Ruffine et al., 2018).

*2.2.4 Central Basin*

The Central Basin is a rhomb-shaped depression having a depth of ~1280 mbsl and area of ~330 km$^2$ (Çağatay and Uçarkuş, 2019) (Figs. 1E, 2C). It is bounded by two boundary faults in the north and south with steep (>10°) continental slopes. The Central High segment splits into branches east of the Cenral Basin, and forms the northern boundary fault the and the 40 m-high scarps of an 8 km-wide inner rhomb-shaped bassinette in the centre (Figs. 1E, 2C). The slopes of the basin are marked by some short (>12 km) canyons.

*2.2.5 Western High*

The Western High between the Central and Tekirdağ basins is ~28 km wide, and has a NE trending folded structure, consisting of more than 6 km-thick folded sedimentary sequence (Bayrakçı et al., 2013) (Fig. 1F). It has a rather rugged morphology with NE-trending folds, with small sagponds and soft sediment deformation and diapiric mud structures (Ruffine et al., 2018; Çağatay and Uçarkuş, 2019) (Fig. 2D). It is cut by the Western Marmara segment of the MMF, forming a deep furrow and fresh-looking scarps (Armijo et al., 2005). This segment forms a right step-over in the west, with the formation of a 1 km-wide and 5 km-long, E-W elongated pull-apart basin (sagpond) with depth of -741 m (Figs 1F, 2D).

*2.2.6 Tekirdağ Basin*

The Tekirdağ Basin is a rhomb-shaped half-graben with an area of ~200 km$^2$ and a maximum depth of -1190 m (Okay et al., 1999; Seeber et al., 2004) (Fig. 1A, 1G). It is filled with sediments tilted southwards towards the active MMF (Fig. 2E). The basin is bounded in the north by a 1.1 km high prominent bathymetric escarpment of the northern boundary fault, dipping south at 11-23° (Okay et al., 1999; Çağatay and Uçarkuş, 2019). The southern slope of the basin is less steep (6-7°) than the northern one (Fig. 1A, 1G). The MMF extends westward from the Western High and forms the southern boundary of the Tekirdağ Basin. In the southwest, the slope is marked by the Şarköy Canyon, which is located on a fault and associated with a submarine landslide (Ergin et al., 2007; Çağatay et al., 2014). The MMF connects with the N-70° striking Ganos segment, which occupies a submarine valley on the western slope of the Tekirdağ Basin (Fig. 1A and 1G) (Okay et al., 1999). The transpressive Ganos segment has caused uplift of the Ganos Mountain (924 m) onshore (Okay et al., 1999) (Fig. 2E). The micro-bathymetry and ROV observations along the southern flank of the Tekirdağ Basin reveal a well-preserved rupture zone that has been assigned to the 1912 Mw=7.4 Şarköy-Mürefte earthquake (Armijo et al., 2005).

## 3. Cores and analytical methods

### 3.1 Cores

Four piston cores (up to 9.45 m-long) and five 0.4 - 1.30 m-long six water/sediment (I) interface cores with undisturbed tops were recovered from the different parts of the SoM on board RVs L'Atalante, Le Suroit and Urania during 2007, 2009 and 2010 cruises (Table 1, Fig. 1A-G). The core sites were selected on the basis of high resolution Chirp seismic profiles (Fig. 2). In the basins, we collected cores from near the depocentres where the turbidite units are the thickest and well defined, thereby providing a possible complete earthquake record. Over the Central and Western highs, sagponds or jogs on the main fault traces were selected for coring. The split cores were described, and then sampled for granulometry and radiocarbon dating. The archive halves were used for MSCL, XRF core scanning and digital X-ray radiography.

We defined coseismic turbidite units (hereafter called turibidite-homogenite: THU) in the cores, based mainly on their sedimentary structures visually observed during the core description and in

X-ray radiographic images and on the μ-XRF elemental profiles. In the SoM, THUs commonly consist of a coarse, often parallel-laminated and occasionally stacked basal layer (TB) and an overlaying sightly normally graded, but homogeneous-looking, mud layer (homogenite, H), which together form a turbidite-homogenite unit (THU; e.g., Beck et al., 2006; Çağatay et al., 2012; McHugh et al., 2014; Yakupoğlu et al. 2019, 2022). The THUs usually have a relatively high gamma and radiographic density and a sharp and erosional basal contact and a transitional (faint) upper contact with the background hemipelagic sediments (HP).

4.2 Laboratory analyses

All non-destructive physical properties and geochemical μ-XRF and granulometric analyses were carried out in ITU-EMCOL Core Analyses Laboratory. Physical properties such as density and magnetic susceptibility were determined using a Geotek Multi-Sensor Core Logger (MSCL) at 0.5 cm resolution. The analyses were carried out according to the standard procedures outlined by (Weaver and Schultheis, 1990).

The cores were analysed for elemental composition at 0.25 mm resolution, using an Itrax μ-XRF Core scanner, equipped with XRF-EDS, X-Ray radiography and RGB colour camera (Croudace et al., 2006; Thomson et al., 2006). A fine-focus Mo X-ray tube was used as the source. The X-ray generator was operated at 30 kV, and a counting time of 20 s was applied. The relative elemental abundances were recorded as counts per second (cps). The cps values of each element were standardized using z-score:

$Z = (x_i - \mu)/\sigma$

where μ and σ are the mean and standard deviation of the population, respectively. The- z-standardized values of the elements  profiles were used as proxies for the source of THUs (e.g., K, Ti, Zr, Fe as siliciclastic input; Ca as biogenic or authigenic carbonate), for redox changes (Mn, Fe), and for grainsize (Zr).

The granulometry was determined in interface cores using a Fritsch laser-diffraction grain-size analyser. The sampling resolution ranged from very 5 cm to a few a few mm within and close to the THUs. Before the analysis, 1 g of wet sample was treated with hydrogen peroxide for 4 hours to remove the organic matter, which was followed by disintegration in 1% calgon solution for 24 hours. The grainsize distribution was determined according to Folk and Ward (1957).

The chronology of the interface cores was constructed mainly by short-lived radionuclide ($^{210}$Pb and $^{137}$Cs) analysis, which was used mainly to date the last earthquake events. The analyses were carried out at the Çekmece Nuclear Research and Training Centre of Turkish Atomic Energy Commission, except core KI07 which was analysed at Hellenic Marine Research Centre. $^{210}$Pb activity was determined by alpha counting of $^{210}$Po. $^{137}$Cs was measured by gamma-ray spectrometry using a germanium detector. Constant Rate of unsupported $^{210}$Pb Supply (CRS) model was used to calculate the SR (Appleby and Oldfield, 1978; Appleby, 2001). For extrapolation of the SR below before last century, downcore compaction was considered, based on the bulk density and dry mass accumulation rates computed using porosity values fom MSCL measurments.

The AMS radiocarbon analysis of benthic foraminifera or bivalve shells collected directly below the base of the THUs was made at NOSAM facility of Woods Hole Oceanographic Institution. Care was taken to collect fresh shells unaffected by reworking and diagenesis. The samples were ultra-sonicated to clean them from mud, washed in distilled water and dried before the analysis. Radicarbon ages were calibrated using the Calib 8.2 program (Stuiver and Reimer, 1993), using the MARINE20 curve (Heaton et al., 2020) with ΔR=-46 and Uncertainty=33 (Siani et al., 2000). The age-depth models of the piston cores were constructed using AMS radiocarbon data, using the R-studio and the script "CLAM" (Blaauw, 2010). The script creates non-Bayesian, cubic spline age-depth model, calculating the %95 Gaussian confidence interval around the best model. Event-free depths, determined from substracting the THUs' thickness, were used in the age-depth model construction. The ages of THUs were determined from the generated depth-age file.

The $\delta^{18}$O and $\delta^{13}$C of carbonate shells and authigenic carbonate nodules were measured using an automated carbonate preparation device (KIEL-III) coupled to a gas-ratio mass spectrometer (Finnigan MAT 252) at the Isotope Geochemistry laboratory of Arizona University. Powdered bulk samples were reacted with dehydrated phosphoric acid under vacuum at 70 °C. The isotope ratio measurement is calibrated based on repeated measurements of NBS-19 and NBS-18. The precision of the method is ±0.1‰ for $\delta^{18}$O and ±0.06‰ for $\delta^{13}$C (1 σ).

## 5. Results

In this section, we describe the lithological and elemental geochemical properties of THU units and their chronology in the studied cores recovered from different morphological regions of the SoM, including the Çınarcık Basin, Central High, Central Basin, Western High and Tekirdağ Basin (Fig. 1, Table 1). Results of previous studies in the İzmit Gulf (Çağatay et al., 2012), Central Basin (McHugh et al., 2014) and Kumburgaz Basin (Yakupuğlu et al., 2019, 2022), located along the Main Marmara Fault are integrated with our results, and discussed under the section 6.

### 5.1 Çınarcık Basin cores

We studied one piston and two interface cores from two locations in the eastern part of the Çınarcık Basin (Figs. 1C, 2A, Table 1). The Interface core KI-08 and piston core KS-10 are located in the depocentre and and interface core KI-07 in the south of the basin. Their lithological description and analyses are reported below.

### 5.1.1 Interface core KI-08

This 71.7 cm long core from the depocentre includes two distinct THUs (Fig. 3A). The upper THU is 16 cm-thick consisting of a 3.5 cm-thick parallel-laminated basal sand-silt layer (TB), and 12.5 cm-thick graded silty clay (homogenite mud, H). The TB part has a sharp erosional contact, and consists of normally graded medium-grained sand (~70%) to coarse-grained silt (~15%) and clay (~15%). The lower THU (THU-KI08-2) is between 63 and 71.7 cm below seafloor (cmbsf) at the core bottom. It consists of a ~2 cm-thick medium-grained sand (TB) and an overlaying 6.5 cm-thick graded sity clay H part at the top. The hemipelagic sediments (HP) are grey green to olive green mud with the uppermost 2.5 cm part having a brownish tint, representing the uppermost oxidized zone. The HP on average consist of 55% silt, 45% clay and and less than 1% sand, with most of the sand fraction consisting of microfossil (mostly benthic foraminifera) shells. The TBs of both THUs are characterized by high gamma density (up to 1.7 cm$^3$/g) and MS (30-35 SI) values and high μ-XRF Ca and Sr counts and low K and Fe. Ti shows enrichment in the THU-1 and depletion in THU-2 (Fig. 3A). The normally graded H parts show upward decreasing gamma density, MS, K, Ca and Fe. The H/HP boundary is marked by changes in the profiles of the elements (e.g.K, Fe and Ca), which is especially the case for THU-2 (Fig. 3A). The HP interval between the two THUs is highly enriched in Mn relative to the THUs. AMS radiocarbon analysis of a marine bivalve shell from 52-54 cmsbsf interval, just below THU-1, in core KI08 provides a calibrated age of AD 1300±60 (Supplementary Table 1).

*5.1.2 Interface core KI07*

This 0.81 cm-long core on the southern margin of the Çınarcık Basin was described and analized for MSCL physical properites and sampled for radionuclide analyses, leaving no archive part for the XRF core scanner and grainsize analyses (Fig. 3B). The core includes three THUs, ranging in thickness from 4 to 7 cm. The upper 12 cm of the core consists of watery, grey-green mud with brownish tint (Fig. 3B). THU-1 between 12-16 cmbsf includes a ~2 cm-thick basal medium-grained sand layer (TB) having a sharp irregular basal contact and a 2 cm-thick homogeneous grey mud (H). THU-2 is 7 cm thick between 41 and 48 cmbsf, with 5 mm-thick dark grey coarse silt lamina (TB) and 6.5 cm grey silty mud (H). THU-3 between 70-74 cm consists of a 3-5 mm-thick fine sand (TB) and 3.5 cm thick silty mud (H). The THUs in KI07 have higher MS and gamma density values than the background sediments with the values reaching to maximum values of 25 SI and 1.65 g/cm$^3$, respectively (Fig. 3B). Gamma spectrometric $^{137}$Cs analysis of core KI07 shows two peaks at 6.5 and 9.5 cmbsf with 24 and 22.5 Bq/kg (Fig. 3C). The vales decay to the negligible values at an extrapolated depth of ~15 cmbsf.

*5.1.3 Piston core KS-10*

This 8.86 m long core near the Çınarcık Basin depocentre includes 30 THUs, with an average thickness of 9 cm and range of 4-29 cm (Fig. 4). The THUs constitute 28 % of the total core succession, and commonly consist of a coarse basal sand-silt part (TB) and a homogenous looking mud (homogenite, H). The TBs are often parallel-laminated with multiple sand-silt laminae, with some lamina showing pinch outs (Fig. 5A, B). Some sandy TBs are disrupted into irregular shapes and detached into lenses (Fig. 5C). Some TBs have sharp and erosional basal contact (e.g. THUs 4, 16, 24). In some TBs, the sand-silt laminae show upward decrease in thickness and grains size, and are often enriched in bivalve shell fragments, ostracod and bentic foraminifera (e.g., *Bulimina* sp., *Brizalina sp.*, *Cibicides lobatula*). In a few TBs, the uppermost lamina is enriched in terrestrial plant material (e.g. THUs 2, 9, 11, 19) (Fig. 5A, B) or contain some 1-2 cm size wood pieces (e.g., THUs 15, 18, 30).

In radiographic images of the thick THUs, the HB passes upward into the H via a normally graded silt layer. The H parts are up to 8 cm in thickness and consists mainly of medium to fine silt. The H has commonly a transitional upper contact with the background HP, which is grey green to olive green homogeneous to faintly laminated silty clay, containing mainly benthic and minor pelagic foraminifera. The HP sediment in upper 17 cm of the core is a watery gray green mud with a brown tint, and below 2.1 m, it contains some brown and black reduced Fe-monosulphide patches.

The TB parts of the THUs show high density in the digital radiographs and high gamma-ray density and magnetic susceptibility in MSCL measurements compared to the H and HP parts (Fig. 4). Some turbidites THUs (e.g. THU-2, THU-21) appear as amalgamated/stacked units, consisting of of two stacked turbisites (Van Daele et al., 2017). A general lowering of magnetic succeptibility occurs below 3.9 mbsf.

µ-XRF elemental profiles of the core shows variable compositon for the THUs (Fig. 4). The TBs are often enriched in Ca and Sr, but depleted in K, Fe and Ti relative to the HP sediments. Calcium profile in the basal parts shows fluctuations paralling the laminated structure (e.g., THUs 3, 4, 9, 12). However, some THUs (e.g., THUs 4, 6, 18, 27) with coarse, quartz-rich TBs are depleted in most elements including Ca, Sr, K, Fe and Ti. The silty to fine sandy TB parts are often enriched in Zr. Basal parts of a few THUs (e.g., THUs 17 and 27) have elevated counts of Ca, Sr, Ti and Fe. Manganese is often enriched in the HP sediments just below the base of THU units.

The age-event free depth (610 cm-long) model of piston core KS-10 is constructed from 14 AMS radiocarbon analysis of fossil shells and plant fragments (Supp. Fig. 1, Supp. Table 1), using the R-studio and the script "CLAM" (Blaauw, 2010). The ages of the THUs based on the age-depth model in the core are presented in Table 2. Acording to the age-depth model, the base of core KS-10 is dated at 9640±160 years BP (7690 BCE).

5.2 Central High cores

We studied one 1.305 m long interface core MEI-01 from the eastern part (-800 m water depth) and one piston core MEG-02 on the fault trace in the western part (-692 m water depth) of the Kumburgaz Basin on the Central High (Figs. 1D, 2B ). In these cores, only one THU near the base of core MEI-01 is observed. Their brief description and chronological anaylsis are important for discussion of the sedimentation rates and site selection of cores for turbidite paleoseismological studies. In the Central High segment part of the SOM, our discussion in section 6.2.3 is based mainly on the long-term sedimentary earthquake records in two ~20 m-long piston cores from the Kumburgaz Basin, recently pulished by Yakupoğlu et al. (2019, 2022) (Fig. 1D).

*5.2.1 Interface core MEI-01*

The main lithology in this core consists of homogeneous green HP mud, which underlays an uppermost 0.8 cm thick, organic-rich dark grey mud and a 6.5 cm-thick thick brown oxidized mud layer below (Fig. 6A). The oxidized zone followed below by a green green HP mud unit, which is underlain by a muddy THU unit between 101-119 cmbsf. The THU consists of a 11 cm-thick sandy silt basal part (TB) and 7 cm-thick normally graded silty H part. The TB part is dark grey to black, parallel-laminated with 1-2 mm thick black sandy silt laminae, containing fine sand- to silt-size shell fragments, which result in high μ-XRF Ca and Sr counts (Fig. 6A). The TB have upward increasing Ca, Sr, Zr and decreasing Mn and Fe profiles. The H part with normal grading shows upward increase in Zr and decrease in Ca, Sr, Ti, Fe, Mn and K,. The gamma density and porosity profiles exhibit general upward increasing trends, with relatively high values in TB part of the THU.

Chronology of core MEI-01 was determined by radinuclide analyses and one AMS radiocarbon dating. [137]Cs profile of the core, shows two peaks at at 3.5 and 9 cmbsf, corresponding to the 1986 Chernobyl and 1963 maximum of atmospheric the nuclear explosion tests (Fig. 6B). Hence, Cs data provides SR rate of 1.52-1.96 mm/yr. [210]Pb profile shows a regular rapid decrease until about 10 cmbsf, and then an slow downward decrease with irregular variation in 12-40 cmbsf interval (Fig. 6C). [210]Pb data provide a SR rate of 1.43 mm/yr, assuming CRS model.

*5.2.2 Piston core MEG-02*

This 3.85 cm long core, located on the fault trace of Central High fault segment, does not include any THUunits, except for a 2 cm-thick fine sand layer between 194.5- 196.5 and a 5 mm-thick sand lens with shell fragents at 190.5 cmbsf (Fig. 7). The core sequence includes two main lithological units: an upper marine unit and a lower lacustrine unit, according to the fossil content (Çağatay et al. 2000, 2015). The marine unit is 1.99 m thick and consists of grey green to dark olive green mud with an euryhaline Mediterranean fauna. The unit includes a dark olive green, locally laminated sapropel unit between 164.5-185.5 cmbsf, with an upper bioturbated layer. The sapropel represents the Holocene main Marmara sapropel (MSAP-1) previously recorded in the SoM cores (Çağatay et al., 2000, 2009, 2015; Tolun et al., 2002). The marine unit passes into lacustrine unit via a 2 cm-thick fine sand layer with shell fragments between 194.5-196.5 cmbsf and 6.5 cm thick rusty brown mud layer between 199-205.5 cmsf. The lacustrine unit consists of grey mud and contains authigenic carbonate nodules with up to 5 cm in diameter at 215, 241 and 272.5 cmbsf, which have $\delta^{13}C$ and $\delta^{18}O$ values which range from -33.2 to -39.5 ‰ VPDB and +1.3‰ and +2.2‰ VPDB (Fig. 7). The unit also includes a dark grey to black Fe-monosulphide layer at 243.5-246.5 cmbsf and spots between 300-386 cmbsf interval.

The density profile shows an upward decrease along the core, with the sandy layers and lenses in the lacustrine unit characterized by relatively high values (Fig. 7). The magnetic susceptibility is relatively high in the lacustrine unit and in the upper 1 m of the marine unit, and gives a high peak (55 msu) in the uppermost part of lacustrine unit. Elements show a general upward decrease in μ-XRF counts (Fig. 7). Fe profile show peaks at the Fe-monosilfide bands and spots, especially in the lacustrine unit. Ca is enriched at the lacustrine/marine transition. K content of the lacustrine unit is higher than that of the marine unit.

The chronology of core MEG-02 is established by the previously determined stratigraphic markers, including the age of the Holocene sapropel (12.33±0.35 and 5.7±0.2 cal), lacustrine/marine transition (12.55 ± 0.35 cal ka BP) (Çağatay et al., 2015). The latter age is further confirmed in the core by an AMS radiocarbon date of 12.65±0.51 cal ka BP at 1.97 mbsf, close to the lacustrine/marine transition. Considering an average sedimention rates of 15 cm/kyr and 30 cm/kyr for the marine and lacustrine units in the core, the core bottom extends back to ~18 ka BP.

5.3 Central Basin cores

Three Marnaut cores, KS-12, KS-13 and KS-18, were previously studied by McHugh et al. (2014) (Fig. 1D). This study only defined the visiually observable coarse basal parts of the THUs and used an age-depth model based on extrapolation of radiocnuclide and AMS radiocarbon data. Here, we provide new high-resolution μ-XRF data to define the H-HP boundaries in piston core KS-18 from the basin depo centre, and then use the chronological data of McHugh et al. (2014) to determine THU ages from the event-free age-depth model based on non-Bayesian statistics.

*5.3.1 Piston core KS-18*

The core is 8.65 m long and includes up to 14.5 cm-thick 15 THUs (Fig. 8). The THU total thickness constitutes 15% of the total core sequence. The THUs with ≥10 cm thickness are THU-2, 10, 11, 15 and 16. The basal TB parts of these THUs consists of ~3 cm-thick parallel-laminated fine sand to sandy silt and up to 9 cm-thick H parts. The fine sand laminae of TBs are 0.5-2 mm-thick and vary from dark gray, black to reddish brown. The sand fraction consists of quartz, benthic framinifera, rare planktonic foraminifera and, bivalve shell and lithic fragments. The HP ediments in the core are commonly yellowish to greyish green hemipelagic mud.

In μ-XRF elemental profiles, the TB parts of theTHUs are enriched in Sr and Ca, and depleted in K, Fe and Ti (Fig. 8). The H/HP boundaries are marked by sharp increases in K, Fe, Sr and Ca in the HP sediments. Common Mn peaks are observed below the base of the THUs. The gamma density and MS profiles do not show any distinct evidence of THU in the core sequence, except for the THU-3 and TH-4 for which both parameters show distinct positive excursions (Fig. 8).

Chronology of core KS-18 is based on the radionuclide and six AMS radiocarbon datings of McHugh et al. (2014) (Supp. Fig. 2, Supp. Table 2). The measurable $^{210}$Pb and $^{137}$Cs concentrations are recorded in the upper 27 cm of the core with highest values of $^{137}$Cs (~175 pCi/kg) at 7.5 and 16 cm (Fig. 8B). The $^{210}$Pb curve apeears to reach background (supported) concentrations (~1.5 dpm/g) at 30-35 cm. The constructed age-depth model based on the radiocarbon datings (Supp Fig. 2) indicates a core bottom age of 6.1 ka BP, and provides the model ages of the THUs listed in Supp. Table 2.

5.4 Western High core

We studied 40.5 cm-long interface core MEI-04 from a ~5 km long pull-apart basin (water depth: -741 m) on the Western High segment of MMF, with the main objective of studying the records of the recent earthquakes (Figs. 1F, 2D) . The uppermost 3 cm thick part of the core consists of brown clayey mud representing the oxic zone (Fig. 9A). The interval between 3-10 cmbsf is green mud with some 0.3-0.5 mm thick biege laminae. The $^{210}$Pb profile, and to a lesser extent the X-ray radiography, show some disturbance between 5-12 cm (Fig. 9A, B). The core below 12 cmbsf is a brownish green mud with two stacked THUs between 19-38 cmbsf. THU-2 consists of two laminated fine sandy silt layers (TB) between 32 and 38 cmbsf and an overlaying 2.5 cm-thick homogeneous mud layer (H). The basal TB in THU-1 has a sharp, erosional basal boundary, and contains ~5 mm-long mud clasts. The overlaying THU-1 has 2.5 cm thick basal sandy mud layer (TB) with shell detritus, overlain by 9-cm-thick silty H (Fig. 9A). The sand fraction ranges up to 8% and the average grainsize varies from 5-15 µm in the lower TB (Fig. 9A). The average grainsize of the overlying HP sediment is 2-3 µm. Both TB layers are enriched in Ca and Sr relative to the HP sediments, but have similar concentrations of K, Fe, Ti and Zr with those in the H parts. The H parts have slightly higher gamma density than the HP sediments and show change in the trend of the elemental profiles, such as Zr and Fe at the boundary with the HP. Manganese shows enrichment just below the base of the THU-1 and in the upper 8 cm of the core with biege laminae (Fig. 9A).

The $^{137}$Cs profile of core MEI-04 peaks at ~3 cmbsf and then show a downward decrease to 6.5 cmbsf, reaching 0.5 mBq/g (Fig. 9C). $^{210}$Pb profile decreases sharply from 140 mBq/g at the core top to 36 mBq/g at 7.5 cmbsf (Fig. 9B). The profile shows some fluctuations between 5-12 cmbsf, and then a regular downward decrease to 22 mBq/g at ~19 cmbsf. $^{137}$Cs and $^{210}$Pb data provide sedimentation rates of 0.65-1.02 mm/yr and 0.62 mm/yr, respectively (see section 6.2.4). One AMS radiocarbon dating on forams, ostracod, and bivalve shells from a sample of at 39 cm below the base of THU provides a calibrated AMS radiocarbon age of AD 470 ± 54 yrs (Suppl. Table S3).

5.5 Tekirdağ Basin cores

Piston core KS-32 and interface core KI-12 from the same depo centre location at -1123 m were studied from the Tekirdağ Basin (Fig. 1G, 2E; Table 1). Detailed sedimentological, geochemical and chronological analyses of the cores are described below.

*5.5.1 Interface core KI-12*

This 88.4 cm long interface core consists of alternating bands of brown, greyish brown and yellow brown mud, with a 5 cm-thick uppermost brown mud band representing the oxic zone (Fig. 10A). The core contains four THUs, which include the stacked THU-2 and THU-3 between 53-72 cmbsf. The THU thicknesses range between 8-18 cm. Except for THU-1, the coarse TB parts of these units are 3 to 4 cm thick, commonly grey sand-bearing coarse silt layers with a sharp basal boundary. The THU-1 is essentialy a mudy turbidite, with a silty basal part with olmost no sand fraction. Each THU is overlain by 5 to 16 cm thick, homogenous silty clay (i.e. H) with usually brown to yellowish brown hue. The TB parts are enriched in benthic forams and silt-size shell fragments.

The basal TB parts of the THU units show high density, MS and µ-XR Ca and Sr counts relative to the homogenite parts and the background sediments, where Fe displays an opposite trend (Fig. 10A). In TBs Ti shows a general enrichment in THUs, except in the THU-1. Manganese is commonly enriched just below the base of THUs (e.g., THU-2), as well as in the upper 19 cm of the core.

The chronology of this core is based on one AMS radiocarbon age of AD 1340±46 from the base of the core (Suppl. Table S1) and radionuclide analyses (Fig. 10B). The $^{210}$Pb profile show some fluctutations in 4-18 cmbsf interval, suggesting some mixing. The $^{210}$Pb values then decrease sharply from 125 mBq/g at 18 cmbsf to 60 mBq/g at 22 cmbsf, and a gradual downward decrease to ~30 mBq/g to 55 cmbsf. $^{210}$Pb data for the core gives an SR rate of 2.2 mm/yr, considering extrapolation to supported levels at 24 cmbsf and using CRS model. The $^{37}$Cs profile show two sharp peaks 3 and 4.5 cmbsf and uniformly high values between 4-18 cm, supporting the physical mixing. $^{137}$Cs profile decrease sharply from 18 cm to 21 cmbsf. No measurements could be made below 21 cmbsf due to instrumental malfuncton. However, considering that the highest peak in the $^{137}$Cs profile at 3 cmbsf represents the 1986 Chernobyl event provides an SR rate of 1.43 mm/yr for the upper part of the core.

*5.5.2 Piston core KS-32*

Core KS-32 includes 20 THUs that are labelled THU-1 to THU-20 from top to bottom, with THUs19 and 20 consisting of two stacked or amalgamated muddy units without an intervening HP (Fig. 11). The THUs in this core range in thickness from 4 cm to 20 cm, with an average of 9.8 cm, and constitute 22.3 % of the core sedimentary succession. The THUs have a sandy to silty coarse TB parts having a sharp basal boundary above the gray green to olive green clayey to fine silty HP sediments. The TBs in some THUs are commonly parallel-laminated sand, sandy silt or coarse silt. THU-5 are THU-8 are the thickest turbidites with 11 and 20 cm thickness, respectively. The basal TB part of these twoTHUs consists of two parts: a basal ~2 cm thick pebbly sand layer containing quartz, other rock-formng silicate minerals and rock fragments, and an overlying commonly parallel-laminated, shell rich coarse to medium sand layer, rich in benthic foraminifera and marine and lacustrine bivalves, including the *Dreissena sp.* derived from the lacustrine unit of the SoM. Some THUs (e.g. THU-19 and THU-20) are muddy turbibites without a visiually distinctive TBs, except for their grey colour, silt-size shell fragments and the accompanied high XR Ca and Sr (Fig.

11). The TB parts of the THUs have high, but upward decreasing μXRF Ca and Sr contents that parallel the change in the shell content and grainsize. In THUs with thick and coarse TBs (e.g. THU-2 and 5), however, Ca and Sr are depleted in the quartz- and silicate-rich basal part parts. Iron, Ti and K are also commonly low especially in the TBs. Manganese shows enrichment with marked peaks in the HP sediments just below and between THUs (Fig. 11). The elemental contents of H parts of THUs and HP sediments are not distincly different, bu show slight changes in the profiles of Ca, Fe, K and Fe at the transition between the two units.

While all THUs throughout the core length are characterized by high gamma density peaks, only those in the upper 3.5 mbsf of the core show distinctly high MS values, where the background MS is is also relatively high (Fig. 11). In the the thick THUs (e.g. THU-2, 3, 5) the gamma density profiles show a gradual upward decreasing trend.

Age-depth model of core KS-32 is based on 19 AMS radiocarbon datings and correlation with the age model of KI-12  (Supp. Fig. S3, Suppl Table S3). The constructed age-depth model indicates a core bottom age of 4.985 (±178) cal yrs BP and core top age of  453±97 cal yrs BP , and provides the model ages of the THUs listed in Table 4. The age-depth model indicates a average earthquake recurrence time of ~240  years for the last ~5.000 years.

## 6. Discussion

6.1 The sedimentological, geochemical and physical properties and origin of the turbidite-homogenite (THU) units

The THUs in the SoM are characterized by a coarse basal sand-silt (TB) and a homogeneous mud (homogenite, H) parts (this study; Beck et al., 2007; Çağatay et al., 2012; McHugh et al., 2014; Yakupoğlu et al., 2019; 2022).  The TB parts of THUs are commonly parallel-laminated and normally graded, and have sharp and often scoured basal boundaries. The laminated part commonly show upward decreasing thickness of lamina and normal grading, together with an upward increasing μ-XRF Ca and Sr counts associated with shell debris (e.g. Fig.  3A). The shell debris consist mainly of benthic forams and bivalve shell fragments that are derived from the shelf edge and upper slope areas via the canyons. Some TBs consisting 2-3 cm thick basal pebbly sand (THU-2 and 5 in core KS-32 in the Tekirdağ Basin, Fig. 11) and contain fresh-brackish water bivalve shell debris derived from the old  lacustrine sediments, while other TBs ( e.g. THU-4 and THU-12 in core KS-10 in the Çınarcık Basin) contain terrestrial plant debris in the upper sand-silt laminae (e.g. THU-2, THU-9, THU-11; Fig. 5 A, B). The pebbly coarse sand in the basal part of TBs are enriched in quartz, heavy minerals (2.8 g/cm$^3$) and in Ti and Fe contents and and commonly have high density and high magnetic susceptibility. The heavy minerals in the sand fraction of THUs (i.e. mainly in TBs) consists of mica, chlorite, opaques, epidote, garnet, amphibole and pyroxene, with some variation between the different basins due to the onland bedrock lithology (Okay and Ergün, 2005). The heavy mineral fractions consists mainly of muscovite, chlorite and opaques in Tekirdağ Basin; amphibole, pyroxene, biotite and chlorite in the Central Basin; and amphibole, pyroxene garnet, epidote, chlorite and opaques in the Çınarcık Basin. The presence of atively high density minerals in the TBs of the Çınarcık Basin THUs are expressed in their higher denstity compared with the THU in the other basins.

TheTBs are overlain by a homogeneous mud layers (homogenite, H) that consist commonly of normal graded, medium to fine silt, with upward increasing clay, K, Ti, Fe contents and decreasing Ca and Sr counts. The Hs have slightly higher density and MS than the HP sediments (Fig. 3, 4). Some THUs have thin (<1-2 cm) TBs and consists essentially mud (e.g. THU-22 and 23 in Fig. 11 in core KS-32 in Tekirdağ Basin; THU in Fig. 6 in core MEI-01 in Kumburgaz Basin; and THUs in cores in the shallow Gölcük Basin in the İzmit Gulf, Çağatay et al., unpublished data). Marked Mn enrichment in the HP sediments just below and between THUs and in the upper part of the cores, represent diagenetic enrichment in a suboxic zone of the sediment column below oxic bottom water conditions (Figs. 3A, 9, 11) (e.g. Thomson et al., 1995; Çağatay et al., 2004 2012; Chaillou et al., 2008). The sharp decrease in MS values below 3.5-4 mbsf in the piston cores is because of the reduction of Fe-oxyhydroxides, below the sulphate/methane interphase in the deep basins (Çağatay et al., 2004; Halbach et al., 2004; Tryon et al., 2010 Drab et al., 2015b; Makaroğlu et al., 2020).

The parallel-laminated stucture of TBs is similar to the Tb division of the Bouma facies (Bouma, 1962) or the T1-T3 division of Stow in the base-cut-out turbidites (Stow and Smillie, 2020). Such lower devisions without a complete turbidite sequence are deposited from a turbididy current when it is more energetic. The upward decreasing thickness and grainsize of the sand-silt parallel laminae in the TBs, together with commonly similar elemental concentrations in the same THU, suggest that the TBs are deposited in the confined SoM basins by waning traction currents, which are reflected or deflected from the opposite slopes, or by resonant water-column oscillatory currents (seiche) (Houghton, 1994; Shiki et al., 2000; Çağatay and Sarı, 2006; McHugh et al., 2006; Çağatay et al., 2012; Yakupoğlu et al., 2022). The upper sand-silt laminae in some coarse TBs contain terrestrial plant debris (e.g., THUs 3, 9, 11, 12, 16, 26 in core KS-10 in the Çınarcık Basin (Figs. 4, 5B), THUs 2 and 5 in core KS-32 in the Tekirdağ Basin (Fig. 11), which strongly suggest deposition from tsunami backwash currents (e.g. Shanmugam, 2011). Presence of sand lenses, balls and disrupted sand beds at some levels (e.g. Fig. 5C) indicate water escape and liquefaction processes.

The overlying H parts of the THUs, on the other hand, are deposited as a bedload from a uniform suspension load following the deposition of the coarse TBs (e.g. Shiki et al., 2000; Beck et al., 2007; Çağatay et al., 2012; Polonia et al., 2017; Drab et al., 2012; Yakupoğlu et al., 2022). Such suspension loads are known to have formed in the water column of the confined basins by remobilization of surficial sediments from the basin slopes and floors, and remain as a thick ponded nepheloid layer for several months before deposition (McHugh et al., 2011, 2016; Ikehara et al., 2016).

The SoM turbidites appear to consists of graded units deposited from mostly single pulses of turbidity flows, rather than from multiple pulses arriving from different canyons and depositing multiple coarse graded amalgamated units (e.g. Gutierrez-Pastor et al., 2013; Ikehara et al., 2016; Goldfinger et al., 2012, 2017). This may be partly due to small size of the SoM basins, with short flow paths of the different canyons, which may hinder the detection of the multiple-pulse units. The stacked THUs with an amalgamated appearance in cores MEI-04 from Western High and core KI-12 in Tekirdağ Basin (Figs. 9,10) are triggered by different events separated by short time interval (see sections 6.2.4. and 6.2.5).

Turbidites can be triggered by various mechanisms including earthquakes, storm waves, hyperpycnal flows, gas hydrate dissociation, sediment overloading, volcanic eruptions and floods. Hence, their use in paleoseismology requires proof of their coseismic origin that can be based on: (i) sedimentary structures and textures (e.g. Prior et al., 1989; Nemec, 1990; Mulder and Syvitski, 1995; Beck et al., 1996, 2007; Chapron et al., 1999; Cita and Aloisi, 2000; Shiki et al., 2000; Schnellmann et al., 2005; Carrillo et al., 2008), (ii) correlation of age-dated turbidites with historical earthquake records (McHugh et al. 2006, 2014; Drab et al., 2012, 2015a; Çağatay et al.,

2012), and (iii) tests such as the synchronicity and confluence tests (Goldfinger et al., 2011, 2017; Patton et al., 2013; Howarth et al., 2021). In this study, we prove the coseismic origin of the THUs in the SoM by using sedimentary structures and textures, correlation with the historical earthquakes, and by using the synchronoicity concept. However, in the SoM, the synchronocity concept would normally apply only for a given individual basin (i.e. fault segment) because of the areally restricted effect of the ground shaking (see discussion below), and the confluence test would not work because of the short flowpath lengths of the canyons in the SoM, (Howarth et al., 2021).

The laminated structure with upward normal grading of laminae, sharp and often erosional basal boundaries, and water esacape and liquifaction structures strongly suggest that the THUs in the SoM are mostly triggered by strong earthquake shaking. The turbidity currents were occasionally violent causing condsiderable erosion on the seafloor. For example, the difference in the radiocarbon age (AD 180) below THU-2 in core KS-32 in the Tekirdağ basin and its modelled age (AD 400) suggests removal of sediment of sediment corresponding to ~200 years (Fig. 11, Supp Table 3). A similar erosive event occurred in the 210 m deep Central Basin of İzmit Gulf during the Mw 7.4 1999 İzmit Earthquake, which removed 5-7 cm of sediment corresponding to 14-20 years (Çağatay et al. 2012).

The triggering of THUs in the SoM by earthquakes is further supported by the correlation of the THUs with historical earthquake records listed in the catalogues (Guidoboni et al., 1994; Ambraseys and Finkel, 1995; Guidoboni and Comastri, 2005; Altınok et al., 2011), as discussed in the next section. It is most likely that in the SoM, only greater than Mw 7 earthquakes with strong-motion parameters (e.g. the peak ground acceleration: PGA within a 40-50 km of the epicentre) can trigger major mass failures and turbidity currents, that could reach the basin depocentres, whereas smaller magnitude and far-field earthquake shakings cause only local turbidity flows. This conclusion is supported by two important observations. First, although turbidity currents were triggered by the Mw 5.8 earthquake of September 26, 2019 in the northeastern corner of the Central Basin (Henry et al., 2022), their sedimentary records were not found in cores in the nearby Kumburgaz Basin and in the Central Basin depocenter (Uçarkuş et al.,in prep). Second, 25 May 1719 earthquake with Ms 7.4 and Ms 6.8 1754 earthquakes with epicenters east of the Izmit Gulf (Ambraseys and Finkel, 1995; Ambraseys, 2002) were not recorded in the sediments of the Karamürsel Basin in the Izmit Gulf (Çağatay et al., 2012).

The THU thickness in the deep SoM subbasin depocentres are broadly similar: 4-29 cm (average: 9 cm) in the Çınarcık Basin, 4-14.5 cm (average: 8.5 cm) in the Central Basin, and 4-20 cm (average: 9.8 cm) in the Tekirdağ Basin. The average recurrence interval in different basins over the last several thousand years range from 222 years in the Tekirdağ Basin to ~300 yrs in the İzmit Gulf (Çağatay et al., 2102) and Çınarcık Basin. The recurrence interval between the two consecutive events in SoM basins is more variable: 90-695 in the İzmit Gulf , 50-930 years in the Çınarcık Basin, 55-770 years in the Central Basin, and 50-780 years in the Tekirdağ Basin. Except in the Tekirdağ Basin, in the other SoM basins' depocentres the THU occurrence frequency show an increase (i.e. event intervals decrease) from the lower (the main Holocene sapropel; 12.6-5.7 ka) to lower sapropel (5.4-2.7 ka) and recent marine (last 2.7 ka) intervals. The average event intervals for these three periods in the Çınarcık Basin are 368, 325 and 232 years (this study), and 287, 114 and 160 years in the Kumburgaz Basin, respectively (Yakupoğlu et al., 2022). The average THU intervals for lower sapropel and recent marine intervals in the Tekirdağ Basin depocentre are 178 and 264 years, respectively, and for the recent marine interval in the Central Basin depocentre is 302. The high temporal and spatial variabilities in occurrence frequency depends on interplay of various factors, including the earthquake magnitude and seismotectonics of the nearby fault segment, the core location in relation to basin morphology and earthquake epicentre, SR, sediment sensitivity to earthquake shaking, and sea level and related paleoceanographic (e.g. salinity) changes (see Yakupoğlu et al., 2022, and references therein).

In particular, water depth and sediment transport routes (i.e. canyons) are important on the THU thickness and frequency in the SoM. The deep basin sites, which have access to sediment input via canyons, include relatively thick and frequent turbidites (e.g., cores KI-08, KS-10 and KS-18). On the other hand, relatively shallower sites near the basin margins and on the intervening highs have thin and less frequent THUs (e.g. KI-07, MEI-04). Another important factor is the cohesivity of sapropelic sediments that appears to be effective in reducing the THU occurrence frequency in the Çınarcık, Central and Kumburgaz basins. In contrast, in the Tekirdağ Basin the the last 2.7 ka period in core KS-32 shows a higher THU occurrence frequency than the upper sapropelic period. This is likely to be due to the long-term changing seismic behaviour of the Western Marmara segment, which is presently creeping with only small (< Mw 4.5) earthquakes (Ergintav et al., 2014; Schmittbuhl et al., 2016; Bohnhoff et al., 2017; Yamamoto et al., 2018), but may have been episodically locked during the earlier periods, producing more frequent Mw>7 earthquakes.

## 6.2 Correlation of sedimentary earthquake records with the historical earthquakes

### 6.2.1 Çınarcık Basin

The $^{137}$Cs peaks at 6.5 cm and 9.5 in core KI-07 represents the 1986 Chernobyl nuclear plant accident and the maximum of global atmosferic nuclear fallout during 1963-1965, respectively. THU-1, observed in 12-16 cmbsf interval below 1965 $^{137}$Cs peak most likely corresponds to the Mw 6.4 1963 earthquake (Fig. 3C). This event is believed to have occurred by normal faulting in south of Çınarcık (Nalbant et al., 1998; Bulut ve Aktar, 2007; Örgülü, 2011), causing tsunami waves with an amplitude of 1 m (Altınok et al., 2011).

The $^{137}$Cs data provides an average SR of 2.26 mm/yr for the upper 9.5 cm cm of the core, coreponding to the period between the year 2007 and 1965. An average compaction of 12.5% relative to the upper 10 cm of the core is estimated from the porosity data for the hemipelagic interval of 16-41 cmbsf between THU-1 and THU-2. With this compaction, and assuming no erosion below Thu-1, a SR of 1.98 mm/yr the 25 cm-thick interval gives an age of 1840 ± 40 yr for THU-2. This age roughly correlates with the Ms 7 1894 earthquake, with the 54-year difference resulting from possible the sediment removal during the event. The 1894 earthquake caused the most damage in the Prince Islands and southern districts of İstanbul, and was accompanied by 1.5 m high tsunami waves (Altınok et al., 2011). THU-3 in core KI-07 occurs 22 cm below THU-2, which corresponds to an age, ~115 year older of than THU-2. Assuming that THU-1 correlates with the 1894 earthquake, the age of THU-3 is estimated to be ~AD 1780. With this estimated age, THU-3 was most likely triggered by the Ms 7.1 May 1766 earthquake in the eastern SoM. The earthquake resulted in devastating casualties and destruction in İstanbul and was accooanpanied by a tsunami, and was followed by the Ms 7.4 earthquake in August 1766 in the west of the SoM (Guidoboni et al., 1994; Ambraseys and Finkel, 1995; Ambraseys, 2002a, b; Pondard et al., 2007).

Based on the SR of 2-3 mm/yr for the deep SoM basins (Beck et al., 2007; Eriş et al., 2008; Drab et al., 2012, 2015a; this study), the AMS $^{14}$C age of AD 1300±60 at 58 cmbsf below the base of THU-1 in core KI-08 is abnormally old, which is likely to be due to the inclusion of reworked shells in the analyzed sample (Fig. 3B, Supp Table 1, Table 1). Considering an average SR of ~2.5 mm/yr, we correlate THU-1 and THU-2 in core KI-08, with THU-2 and THU-3 in core KI-07. This THU-matching for the interface cores is supported by their similar correlation with the THUs in piston core KS-10, which was recovered from the same location as KI-08, and dated by age-depth modelling (Table 2, Fig. 12, Fig. S1). The model ages of THU-1 and THU-2 in core KS-10 are AD 1889±123 and 1789±88, which indicates their association with the 1894 and 1766 earthquakes. Another supporting indirect evidence for the THU correlation in our Çınarcık Basin cores is the presence of similar turbidite records of the 1894 and 1766 earthqukes in other Çınarcık basin cores (Drab et al., 2015a; Fig. 1C).

The Mw 6.4 1963 earthquake event appears to be the youngest earthquake record in the Çınarcık Basin, which is recorded only in core KI-07, but absent in the Çınarcık depocentre cores KI-08 and KS-10, probably because of the relatively small magnitude of the event and the distant location of the two cores from the epicentre. Interestingly, the Çınarcık basin cores did not record the 1999 Mw7.4 İzmit earthquake, indicating that the sedimentary records of the earthquake are restricted to the İzmit Gulf basins. The greater thickness of the THUs in core KI-08 compared with those of the corresponding THUs in core KI-07 can be explianed by the depocentral and epicentral position (i.e. Prince Islands segment of the MMF) of core KI-08 (Ambraseys, 2002b).

In core KS-10, THU-1 triggered by the 1894 earthquake consists of a 5 cm-thick TB and an overlaying 3 cm-thick silty mud (H). The TB part is composed of normally graded two fine- to medium-grained sand layers, sparated by 1.5 cm thick silt layer. THU-2 assigned to the 1766 earthquake is represented by a stacked THU in the 40-69 cmbsf interval (Fig. 4). The lower part consists of seven stacked sand laminae that shows a general upward decreasing thickness and grain size, deposited by a single pulse (Fig. 5A). The middle thick sand lamina includes terrestrial plant fragments in the upper part, representing the tsunami-backwash material (Fig. 5A), which supports the tsunami occurrence recorded in the historical records following this event (Guidoboni et al., 1994; Ambraseys and Finkel, 1995). The 3 cm-tick fine sand layer and nd 4 cm-thick silty H layer forming the upper part of the stacked package was likely formed during the 7.4 August 1766 earthquake in the western SoM.

The THU-3 in core KS-10 is 12 cm-thick between 73-85 cmbsf (Fig. 4). With a model age of AD 1541±78, this THU is correlated with the 1509 İstanbul (Constaninople) earthquake with M=7.2 ± 0.3, which is also known as the Little Apocalypse (Ambraseys, 2002b; Guidoboni et al., 1994). This earthquake was associated with an important tsunami on the İstanbul coasts of the SoM (Altınok et al., 2011). No terrestrial plant fragments was found in the THU. However, evidence for the tsunami is represented by 1.5-3 cm-thick three sand layers overlaying an erosional basal contact and showing upward decreasing thickness and grainsize (Fig. 4). The record of the 1509 İstanbul earthquake is found also in other Çınarcık cores by Drab et al. (2015a) as well as a relatively weak record in the İzmit Gulf cores by Çağatay et al. (2003, 2012) and McHugh et al. (2006).

THU-4, which consists of 1-2 cm-thick three very fine sand to coarse silt layers (TB) and an overlying 4 cm-thick silty H, is dated at AD 1373±106 according to the age-depth model (Table 2). With this age range, it most likely correlates with the M=7 October and November 1343 earthquakes, which caused widespread damage in İstanbul (Constantinople) and settlements in central and western SoM coastline and resulted in a tsunami (Guidoboni et al., 1994; Ambraseys, 2002 a, b). The record of this earthquake was also observed in cores from the north of northern Çınarcık Basin by Drab et al. (2015a). Well documented historical records indicate a sequence of earthquakes which lasted for a year. In consatinople, part of the Theodosian walls collapsed and palaces, churches including the east side of the St.Sophia church was also damaged (Guidoboni et al., 1994).  Although, a large tsunami was reported in the Bosphorus with dragged boats and  mud and dead fish on land, THU-4 does not show a clear evidence tsunami.

THU-5 is a 16 cm-thick major event deposit, which consists of  5 cm-thick TB part with three medium to coarse sand layers interbedded with silt layers and a 11 cm-thick silty H part. The sand layers contain shell debris, and show liquefaction and water escape structures with sand injection and formation of sand lenses (Fig. 5C). THU-5, with its model age  AD 737±96 (Table 2),  most likely represents the record of Ms 7.1 (I = IX-XI) 26th October 740 earthquake with an epicentre located south of İstanbul (Soysal et al., 1981; Ambraseys and Finkel, 1995; Ambraseys, 2002a,b; Guidoboni and Comastri, 2005; Altınok et al, 2011). Sedimentary records of this earthquake was previously observed in the various sub-basins of the İzmit Gulf (McHugh et al., 2006; Çağatay et al. (2012) as well as in the Çınarcık Basin (Drab et al., 2015a). This earthquake was reported to have caused devastating destruction in İstanbul and İzmit, and resulted in a tsunami along the coasts of the İzmit Gulf (Altınok and Ersoy, 2000; Altınok et al., 2011).  Another earthquake that took place within time range of  the model age for THU-THU-5 is the AD 715 (I = IX) eartquake, which according to Pınar ve Lahn (2001), caused widespread demage in İstanbul and İzmit, but this event lacks detailed information in most historical records and earthquake catalogues.

THU-6 has a model age of AD 527±98 (Table 2). Within the age range, the most likely event corresponding to this record is the 14 December 557 earthquake (I=IX; M: 7.0), with an epicentre estimated to be between İstanbul and İzmit (Altınok et al., 2011, and references there in). Many houses and churches were destroyed, particularly in the coastal district of Küçükçekmece (Rhegion), southwest of İstanbul. The sedimentary record of this event consists of a 5 cm thick very coarse, laminated basal sand with shell debris (TB) and and 4 cm-thick silty mud (H). A similar sedimentary record of similar age has been reported by Sarı and Çağatay (2006) in another core from the Çınarcık Basin (Fig. 1C).

THU-7 in core KS-10 consists of a laminated a 3.5 cm-thick, brown, laminated medium- to coarse -grained sand (TB) and a 5 cm-thick H (TB). The upper 1 cm-thick organic-rich dark gray silt lamina in the upper part of the TB. contains terrestrial plant detritus. With a model age of AD 437±132, THU-7 can be associated with the M:6.6 (I: VII–VIII) AD 407 Çınarcık Basin earthquake, which had a devastating damage in İstanbul (Constaninopolis) according to Byzantian historian Marcellinus (Table 2) (Guidoboni et al., 1994). It was associated with a tsunami with a large number number of corpses washed on the shoreline in Bakırköy (Hebdemon), west of İstanbul. The presence of terrestrial organic matter in THU-7 supports the tsunami occurrence and correlation with the event. Within the age range of THU-7, another candidate is the AD 358 İzmit earthquake (M=7.4, I=IX), which caused was very large scale destruction around the İzmit Gulf killing 30,000 people and resulting in a landslide associated tsunami (Soysal et al., 1981; Guidoboni et al., 1994; Altınok and Ersoy, 2000; Ambraseys, 2002a, b).

The remaining earthquakes records of THU-8 to THU-30 in the lower part of core KS-10 occurred during classical antiquity and prehistorical period according to the age-depth model. The historical records of earthquakes for the period BCE are less frequent and most likely to be less reliable than the AD events, especially in terms of epicentral locations. Epicentre of earthquakes during these periods are assigned according to the existing settlements at the time. It is important to note that İstanbul (Constaniople) was not yet established at these times, but instead other Thracian settlements on or near the central and western coasts of SoM, such as Perinthus (Marmara Ereğlisi), Hellespont (Çanakkale), Lysimachia near the Gulf of Saros and other towns on the Thracian Chersonese (Gallipoli Peninsula), which are located 50-120 km from the Çınarcık Basin. Hence, the lack of settlements in the eastern Marmara region hinders unambigious correlation of the Çınarcık Basin sedimentary records with historical earthquake records during the BCE period.

THU-8 and THU-9 have overlapping model ages of 303±122 BC and 359±118 BC, respectively (Table 2). With its laminated, normally graded sand structure and terrestrial plant detritus content, THU-9 appears to represent a major earthquake event and evidence of a tsunami in the SoM. Within the age ranges, the 287 BC and the 360 BCE earthquakes are the likely canditates to trigger THU-8 and THU-9. However, according to the reports of Demosthenes (384-322 BC) and Aristotle (384-322 BC), these earthquakes caused the most damage in the ancient city of Lysimachia (Mürefte-Şarköy) and the Hellespont (Çanakkale) in Gelibolu Peninsula, (Guidoboni et al., 1994). The city of Lysimachia was completely ruined after it was founded by king Lysimachus twenty-two years earlier. The sedimentary record suggest that the effect of the earthquake with an epicentre in the western SoM might have extended all the way to the Çınarcık Basin. Alternativley, the lack of settlements and hence the absence of historical eartquake records in the eastern Marmara region might have hindered its documentation in the east, hence proper assignment of the epicentre.

THU-10 in core KS-10, with 13 cm thickness and evidence of a tsunami backwash detritus, is represents a major earthquake (Fig. 4). With a model age of 586±82 BC, it is tentatively linked to the late December of 427 BC Perinthus earthquake, which is the only and oldest historical event recorded for the region (Guidoboni et al., 1994). Hippocrates (~460 - 370 BC) in *Epidemics* mentions an earthquake at Perinthus (presentday Marmara Ereğlisi) shortly after the winter solistice (Guidoboni et al., 1994). Perinthus was the main settlement in the region at the time. This event affected large areas extending to the Black Sea coast NE of İstanbul, and its sedimentary record was aslo found in the Gulf of İzmit (Çağatay et al., 2012).

The base of core KS-10 is dated at 9,644±162 years BP (7694 BC), which provides an average earthquake repeat time of ~320 years. However, the interval between two consecutive THUs (earthquakes) events are highly variable, ranging from ~ 50 years to 660 years.

### 6.2.2 Central High

Core MEI-01 was recovered from -800 m water depth in the Kumburgaz Basin to date the last earthquake event on the Central High segment, which is represented by the 18 cm-thick muddy THU (Figs. 1D, 2B, 6A). The $^{137}$Cs profile of core MEI-01 provides SR of 1.52- 1.96 for the last

46 years (since 2009), considering the peaks of the 1986 Chernobyl and the 1963 maximum of satmospheric uclear explosions test (Fig. 6B), wheras [210]Pb data gives a SR of 1.43, assuming the CRS model. These rates correspond to 1.32-1.65 mm/ (average: 1.49 mm/yr) with a cumulative compaction of 8% for the upper 101 cm of the pelagic sediments above the THU in the core. Considering these SRs, the THU is dated between AD 1245 and 1395.

With this age range, we tentatively correlate the THU event in core MEI-01 with the M7 AD 1343 Marmara earthquake (Intensity IX, ) that affected the central northern shores of SoM and İstanbul and was accompanied by a tsunami (Guidoboni and Comastri, 2005; Ambraseys, 2002 a, b). Other possible candidates are 1354 and 1437 earthquakes, but these appear to have affected the Gelibolu and Eceabat, with their epicentres in the western SoM (Ambraseys, 2002; Yaltırak et al., 2003; Guidoboni and Comastri, 2005).

The 12-40 cmbsf interval in the core, with the [210]Pb fluctutations, includes faintly laminated sediments without any sign of disturbance or mixing (Fig. 6A, C). Hence, these enigmatic fluctuations cannot be related to earthquake shaking or bioturbation,. Rather, they are likely related Instead to changes in the rate of erosion or atmosphic precipitation in the drainage area, which includes the Büyükçekmece lagoon, fluvial and canyon systems.

Although core MEI-01 does not show the records of the more recent the Mw>7 1766 and 1509 earthqukaes, these records are found in an interface core from the depocentre (-~850 m depth) of the same basin by Uçarkuş et al. (in prep.). The absence of the same records in core MEI-01 is most likely due to the core location, ~ 50 m higher than the depocentre at the eastern edge of basin. The same is true for the piston core MEG-02 extending back to ~18 ka BP; this core does not include any THUs despite being located on the main Central High fault segment west of the Kumburgaz basin, because the location is ~160 m higher than the sediment depocentre (Fig. 1D).

However, core MEG-02 includes some authigenic carbonate nodules in the lacustrine unit, with $\delta^{13}$C values of -33.2 to -39.5 ‰ VPDB, which indicate a dissolved carbon source derived from anaeorobic oxidation of thermal methane (Devol et al., 1984; Boetius et al., 2000; Niemann et al., 2006; Crémière et al., 2012; Çağatay et al. 2018) (Fig. 8). These nodules were formed within the sediments near the seafloor by anaerobic oxidation of thermogenic methane that was emittted along the MMF fault (e.g. see Ruffine et al., 2018 and references there in). The methane emission from the fault might have been triggered by different processes including seismo-tectonic activities and gas hydrate destabilization due to climatic warming of the Mediterrnaenan waters (Ménot and Bard, 2010; Crémière et al., 2013; Çağatay et al., 2018; Teichert et al., 2018).

The long-term earthquake records in the Kumburgaz Basin from up to 21m-long two giant piston cores were published by Yakupoğlu et al. (2019, 2022). These records include 28 THUs deposited between 813 cal. a. BP and ~6116 cal. a. BP, nine of which are tentatively correlated major historical earthquakes (Yakupoğlu et al., 2019). These include AD 989, AD 869 or AD 862, AD 740, AD 557, AD 447 or AD 478, AD 407, AD 120–128 or AD 180–181, AD 29 or AD 69, and 287 BC or 427 BC. The most recent important earthquakes of AD 1766 and 1509 were not detected in the piston core, because the top ~2.5 m, corresponding to the last ~810 yrs, was not recovered. The core provided an average THU recurrence interval of 220 years over the last 6 kyrs, with the interval between the consecutive events ranging between 40 yrs. and 425 yrs.

### 6.2.3 Central Basin

In core KS-18, 16 THUs are dated by construction of non-Bayesian cubic spline age-depth models using radionuclide and six AMS radiocarbon data of McHugh et al. (2014) (Fig 8, Supp Fig. S2, Table 3, Table S2). The high concentrations of $^{137}$Cs and unsupported $^{210}$Pb at the core top indicate that the core has recovered recent sediments younger than the last 100 years (Fig. 8). The occurrence of the measurable $^{137}$Cs down to the limit of measurable unsupported (excess) $^{210}$Pb at 30 cm suggest sediment mixing by bioturbation. The unsupported $^{210}$Pb curve suggest an extrapolated core depth of 30-35 cmbsf for reaching the supported $^{210}$Pb value, which corresponds to an age of ~110 years (5 x half-life of $^{210}$Pb) before year 2007 (date of the core collection) and 53 years before 1950 (see e.g. Appleby and Oldfield, 1978; Appleby, 2001). The disturbance in 14-25 cmbsf interval is therefore younger than the year 1900. Similar disturbed layers are also observed in intervals 13-17.5 cmbsf and 9-16.5 cmbsf of cores KS-12 and KS-13 in the basin studied by McHugh et al. (2014). This disturbed layers are likely to be due to bioturbation rather than to the M6.4 1963 Çınarcık or M7 1964 Manyas earthquakes, as proposed by (Mchugh et al., 2014), considering that the epicentres of these events are ~50 and ~70 km away from the Central Basin, respectively.

The most recent THU-1 in core KS-18 is dated at AD 691±94 (Table 3), and thus, correlates well with the 26th October 740 earthquake with Ms = 7.1 and I = IX-XI with an estimated epicentre south of İstanbul (see section 6.1). The THU characterizing this event consists a 5.5 cm-thick TB and 9 cm thick muddy homogenite (H). The TB has a sharp basal boundary, and consists of grey to reddish brown laminated fine sand (Fig. 8). The same event is also recorded in cores KS12 (Fig. 1E) (McHugh et al., 2014). THU-2 and THU-3 have average ages of AD 300 and AD 233 with a large error margins of 167 and 191 years (Table 3). Within the wide age range, the possible triggering events for the deposition of these THUs are the August 358 (M=7.4, I= IX), October 268 (I=VIII), 402, 407 (I:VII-VIII) and 447 (I: IX–XI ) earthquakes. However, the first two earthquakes have their epicentre in the İzmit Gulf (Altınok et al., 2011), while the fifth century earthquakes are reported to have caused geat damage in Istanbul and the Marmara coasts in general, with tsunami flooding of the Bakırköy (SW of İstanbul) and Bosphorus coastline (Guidoboni, 1994; Abraseys, 2002; Altınok et al. 2011).

In addition to the above earthquake records, McHugh et al. (2014) reported the records of 1766, 1509, 1343, 860 and 557 earthquakes in the Central Basin cores (Fig. 1E). However, the records of the 7.4 1766 and August M 7.2 1509 earthquakes are present only in core KS-13 (McHugh et al., 2014). This suggest that core KS-13, with its 23 THUs and location near the foot of the 40 m-high fault scarp, appears to be a sensitive recorder of mass flows triggered by earthquake shaking (Fig. 1E).

The KS piston cores in the Central basin does not include the record of Mw 7.4 1912 Şarköy-Mürefte earthquake (McHugh et al., 2014 and this study), whereas Drab et al. (2012) report its presence in core Klg-02 in the Central Basin (Fig. 1E). However, the evidence including the X-ray radiography image and grainsize data for this event in core Klg-02, is not convincing, and appears to be due to bioturbation or an artefact of piston coring. Drab et al. (2012) also report other 10 THUs, and correlate the recent six to 1766, 1354 or 1343, 1063, 557 or 437 earthquake events.

Although our results from core KS-12 suggests that the ground shaking by the 1912 earthquake did not reach the threshold to trigger a THU in the Central Basin, Armijo et al. (2005) and Aksoy et al. (2010, 2021) claim that the Mw 7 1912 earthquake ruptured the full length of the western segment all the way to the Central Basin. This claim is based on the fresh submarine scarps on the Western High (Armijo et al., 2005) and the estimated rupture length from onland paleoseismological studies (Rockwell et al. 2009; Altunel et al. 2004; Aksoy et al. 2010, 2021).

*6.2.4 Western High*

Interface core MEI-04 from the 5 km-wide transtentional basin on the Western High includes two stacked THUs in 18.5-38 cmbsf and a disturbed sediment layer in 5-12 cmbsf interval (Figs. 1E, 2D, 9A, 2D). The timing of these events can be estimated from the radiocnulide data (Fig. 9B, C). The $^{137}$Cs peak at ~3 mbsf at most likely represents the year 1963 when the atmospheric nuclear tests reached a maximum, while stable background value of 0.5 mBq/g at 6 cmbsf represents the year 1950 when nuclear exlosions started (Fig. 9B) (Appleby and Oldfield, 1978; Appleby, 2001). Thus, the $^{137}$Cs data provide SR rate of 0.65-1.02 mm/yr, while the $^{210}$Pb data indicate a SR rate of 0.62 mm/yr, assuming CRS model. Using these data age of the stacked THUs 1 and 2 at ~19 cmbsf is bracketed between AD 1700 and 1820, which corresponds well to the M 7.4 August 1766 earthquake in western SoM (Guidoboni et al., 1994; Ambraseys ve Finkel, 1995; Ambraseys, 2002a, b; Pondard et al., 2007). The stacked THUs were likely triggered consecutively by strong ground motion of the August 1766 earthquake and its strong aftershock, or alternatively by the May and August 1766 earthquakes.

Compared to the radionuclide data, the AMS radiocarbon age of AD 470 ± 54 at 39 cm below the base of THU appears to be too old. This overly old $^{14}$C age may be due to the possible presence of some reworked shells in the dated sample and/or some sediment removal below the base of the THU.

Considering the radionuclide data, the disturbed sediment layer at 5-12 cmbsf is dated to be older than 1950 and younger than AD 1900. With this age range, this layer is likely to have been caused the Mw7.4 1912 Şarköy-Mürefte earthquake. A similar muddy THU record of the the same earthquake was found by Drab et al. (2012) in cores Klg06 and Klg07 on the western and eastern edges of the Central High (Fig. 1E).

The record of the 1766 earthquake was found also by Drab et al. (2012) in piston cores Klg06 and Klg07 on the Central High (Fig. 1E). Moreover, using correlation with the Tekirdağ Basin Klg cores, Drab et al. (2012) report also the muddy THU records of the 1354 or 1343, 1063, 557 or 437 earthquakes on the Central High cores. However, all the THUs in the Klg cores, including that of the 1766 earthquake, are less distinct muddy turbidites, compared to the well defined THU with a distinct sandy TBs in core MEI-04.

*6.2.5 Tekirdağ Basin*

The age estimates of the four recent THUs in core KI-12 are based on radionuclide data and one AMS radiocarbon age (AD 1340±46) near the core bottom (Figs. 10, Table S3). The $^{210}$Pb and $^{137}$Cs profiles show uniform and fluctuating high values in the upper 18 cm of the core, strongly suggesting physical mixing by bioturbation, earthquake shaking or core handling (Fig. 10B). Using the SR rates of 1.43 mm/yr and 2.2 mm/yr derived from the $^{137}$Cs and $^{137}$Pb data, the age of THU- is estimated beween AD 1880 and 1925. This age range closely matches with the Mw 7.4 1912 Şarköy-Mürefte earthquake. Similar records of the 1912 earthquake were also found by McHugh et al. (2006) and Drab et al. (2012) in the Tekirdağ Basin.

Using a SR of 1.26 and 1.93 mm/yr with 12% compaction for the 34-52 cmbsf interval and assuming no erosion below the base of THU-1, duration for the depositon of the18 cm-thick hemipalagic sediment unit between THUs 1 and 2 is ~ 143- 93 years, which provides an age range of AD ~1769 -1818 for the stacked THU-2 and THU-3 in core KI-12. The only well-known earthquake occurring in the area after the 1912 earthquake is the August 1766 event. Considering their stratigraphic position and similarity with the stacked THUs in core MEI-04 on the western Central High, we associate THUs 2 and 3 in core KI-12 with the 1766 earthquakes. In that case, some time gap of up to 50 years between the estimated and assumed ages suggests removal of 6-10 cm-thick sediment from the sea floor during the 1912 earthquake.

The age THU-4 in core KI-12 is provided by the AMS radiocarbon age of AD 1347 (range: AD1280 - 1419) (610±46 cal yr BP), directly measured on a sample just below the base of THU-4 (Table 4). The most lileky earthquake correponding to this age is the November 6, 1344 earthquake that occurred in Western SoM, as part of an earthquake sequence that occurred from autumn of 1343 to the summer of 1344 and affected the Constantinople and the Thracian coast. The strongest shock in this series occurred on 6 November 1344, destroying the fortresses in Gazikoy and citadel in Hoşköy, west of Tekirdağ, and also collapsed some walls of the St. Sophia church in Constantinople (İstanbul), that was already damaged by the 1343 earthquake (Guidoboni and Comastri, 2002). MacHugh et al. (2006) and Drab et al. (2012) found a similar muddy THU in their Tekirdağ Basin cores, having a similar lithology and age range to THU-4 in core KI-12 (Fig. 1G), and tentatively assigned to the 1354 or 1343 earthquakes.

The age model for core KS-32, based on 11 AMS radiocabon datings, indicates that the upper part of the core correponding to the last 453±97 cal years is not recovered (Supp. Fig. S3, Table 4).

Hence, the four THUs observed in core KI-12 are missing in this core. The first THU at 73-78 cm interval in this core is dated AD 968 (range:AD 881-1062), which may match with AD 989 and AD 945 earthquakes (Guidoboni et al. 1994). However, 26 November AD 989 earthquake appears to be a more likely candidate, because it affected mainly the western Thrace region and also recorded in Constantinople, whereras the latter earthquake (I=VI) affected only Constantinople.

The THU-2 in core KS-32, with a 5-cm thick, pebbly sand in its basal part including shell fragments from lacustrine sediments, was likely triggered by by a strong earthquke shaking associated by a tsunami. Cossidering radiocarbon age 180±54 AD just below , this THU's modeled age of AD 400 (range: 307-487) suggest erosion of some sediment thickness by corresponding to ~ 200 years period, and indicates strong ground motion. The most likely earthquake that could have triggered THU-2 is the 26 January 447 earthquake that affected a large area from İstanbul and İzmit in the east to the Gelibolu- Dardanelles (Hellespont) region in the west, ruining many cities and killing thousands of people (Guidoboni et al. 1994). This great earthquake ocurred during the reign of Theodosius II, and triggered a tsunami that submerged the Marmara islands and sank many ships. The paleoseismological studies by McHugh et al (2006) and Drab et al. (2012) in the Tekirdağ Basin report a THU record that they assign to the 557 or 437 in the Tekirdağ Basin. However, the 437 earthquake, with a relatively low magnitude (Ms=6.8; Ambraseys, 2002) and its epicentre in eastern SoM, is not likely to cause a THU in the Tekirdağ Basin.

THU-3 at 147 cm, with a model age AD 230 (range: AD 133-331), can be tentatively assigned to Ms AD 180 or AD 268-270 earthquakes (Guidoboni et al.1994; Ambraseys, 2002b). However, historical reports relate these earthquakes with an epicentre in İzmit (Nicomedia).

THU-4 has a model age of 153 (range: AD 32- 90 BC). The historical records become scanty and unreliable around and pre-BC period. One possible the event is as the mid-first century the Dardanelles (Hellespont) earthquake that damaged settlements the shores of the Galibolu Peninsula (Guidoboni et al., 1994). Another event within the given age range is the AD 120/128 earthquake that caused damage in Biga (Parium), Erdek (Cyzicus)-İzmit (Nicomedia)-İznik (Nicea), and was associated with a tsunanmi (Guidoboni et al., 1994). Emperor Hadrian gave generously from the public purse to restore the Nicomedia (İzmit).

THU-5 at 222 cm core depth is another sedimentary record in core KS-32, which is most likely associated with a strong earthquake and a tsunami. Its has a model age of 217 BC (range: 100-

335 BC). This THU likely correlates with the strong 287 BC earthquake, which caused widespread damage in in Dardanelles region (Hellespont) including the Gelibolu Peninsula (Thracian Chersonese) and Mürefte-Şarköy (Lysimachia) (Guidoboni et al., 1994). The city of Lysimachia, founded by king Lysimachus twenty-two years earlier, which completely destroyed. With the remainig 15 THUs, the KS cores provides an average recurrence time of 222 yrs for the last 5000 years). However, the interval between two consecutive THUs range from 52 yrs to 668 yrs.

*6.2.6 İzmit Gulf*

Earthquake records of the İzmit Gulf segment were obtained from cores in the Central (Karamürsel) and Gölcük Basins by Çağatay et al. (2012) and in Western (Darica) Basin by McHugh et al. (2006) (Fig. 1G). The THU in piston core MARM05-115 from the 210 m-deep depocentre of the Central Basin has nine THUs, including the uppermost THU triggered by the Mw 7.4 17 August 1999 İzmit earthquake (Fig. 13). In the interface cores IZ-112 and IZ-115, the THU correlated with the 1999 earthquake has a total thickness of 32.5 cm and consists of three parts above erosional basal bounday: a graded coarse sandy-silty layer, an intermediate laminated silt layer and a homogeneous mud layer (Çağatay et al., 2012). Some THUs (e.g. THU-7 and THU-5) show liquefaction structures such as irregular lenses and balls, which are likely produced by seismic shaking.

The THU-2 below is corelated with Ms>7.4 1509 earthquake ( Ambraseys and Finkel, 1991, 1995; Altınok and Ersoy, 2000; Ambraseys, 2002, Fig. 13). The 1509 earthquake caused damage to the Hersek mosque in west of the Kumburgaz Basin according to the historical records (Witter et al., 2000). In addition to the Kumburgaz Basin core, the records of 1509 earthquake were also found in sediment cores from the Darıca Basin in western İzmit Gulf and the Hersek Lagoon (Çağatay et al., 2003; McHugh et al., 2006; Bertrand et al., 2011) (Fig. 14).

THU-3, with a calibrated radiocarbon age of 1272–1387 AD, is correlated with the 1 June 1296 (intensity, I=VII) (Ambraseys, 2002; Altınok et al., 2011) (Fig. 13). THU-4 is dated between 730 and 959 AD range. Three large earthquake events occurred during this interval in the Marmara region within the age range of this THU unit: 862, 865, and 875 AD (Soysal et al., 1981; Guidoboni et al., 1994). According to historical reports, the 865 AD event is the most likely one affecting the İzmit area most. THU-5 has an age range of 645–735 AD, which matches with the of Ms 7.1 (I= IX–XI) 26 October 740 AD earthquake (Soysal et al., 1981; Ambraseys and Finkel, 1995;

Ambraseys, 2002; Guidoboni and Comastri, 2005; Altınok et al., 2011) (Fig. 13). Most historical accounts place the epicentre of this earthquake south of İstanbul and reports wide scale destruction in İstanbul and around İzmit Gulf, with an associated a tsunami. Two pieces of evidence support the correlation of THU-5 with the 740 AD event. First, a similar record of this earthquake is also found by McHugh et al. (2006) in the Western Basin of the İzmit Gulf (Fig.1B; Fig. 14). Second, a relatively thick (5.1 cm) coarse sandy basal layer of THU-5, with a laminated bi-directional cross bedding structure, which strongly suggest deposition from reflecting turbidity currents or the effects of seiche like water column oscillations associated with a tsunami event (Çağatay et al., 2012).

THU-6 and THU-7 have overlapping calibrated ages of 111-594 AD and 150-391 AD (Fig. 13). 24 August 358 earthquake (M=7.4, I= IX) and October 268 AD earthquake (I=VIII) (Altınok et al., 2011 and references therein) are possible earthquakes within the age range which might have triggered the THUs. Both these events ruined the settlements and killed thousands of people around the İzmit Gulf. THU-8 has an age range of 711–176 BC. The only earthquake event within this age range is the 427 BC earthquake, with an estimated epicentre location close to Marmara Ereğlisi (Perinthos) (Guidoboni et al., 1994). This event is reported to have affected large areas extending to the Black Sea coast northeast of İstanbul.

Interestingly, some earthquakes with epicentres assigned to the İzmit Gulf area are absent in the cores from the Central Basin. These include the is the Ms 7.4 May 25 1719 earthquake east of the Hersek peninsula in the Gulf of İzmit (Ambraseys, 2002, Ambraseys and Finkel, 1995). The absence of its record in the studied core suggests that its epicentre is probably located east of the İzmit Gulf, rather than being within the Gulf. The record of Ms 6.8 September 2, 1754 and Ms 7.3 July 10, 1894 earthquakes with epicentres in the Çınarcık Basin (Ambraseys, 2002) was also not found in cores of the İzmit Central Basin, but observed in the Western Basin of the İzmit Gulf (McHugh et al., 2006; Fig. 14). This likely indicates that the ruptures of ground shaking by these two earthquakes did not extend into the Karamürsel Basin (Çağatay et al. 2012). Moreover, some earthquakes epicentres assigned to the south of İstanbul in the Çınarcık Basin (e.g., AD 865 earthquake) may be incorrect, and that the effect of other major events (e.g., 740 AD and 1509 earthquakes) may have extended to the İzmit Gulf. Assigning most earthquake epicentres close to İstanbul is understandable, considering that it has been the main city in the region for the last 2000 years, and thus, has been affected by large earthquakes with epicentres located elsewhere in the region. This may also be the case for the 427 BC earthquake; with its epicentre assigned to near Perinthus, a major settlement 100 km west of İstanbul at the time, although this earthquake might have occurred in Çınarcık or even the İzmit Gulf.

## 7. Summary and conclusions

We have investigated the coseismic turbidites (turbidite-homogenite units: THU) in sediment cores on different segments of the Main Marmara Fault crossing the basins and highs in the northern SoM (Fig. 14). The cores were analysed using X-ray radiography, grainsize, MSCL physical properties and μ-XRF elemental analyses, and age-dated by radionuclide and AMS radiocarbon analyses. Using the correlation with the historical earthquakes in different cores and integrating the previous submarine paleoseismological data (Sarı and Çağatay, 2006; Çağatay et al., 2012; McHugh et al., 2006, 2014; Drab et al. 2012, 2015a), we tested the coseismic origin of the THUs in the SoM using the synchronoicity concept of Goldfinger et al (2017).

Coseismic turbidites (THU) observed in the Sea of Marmara cores consist of a graded and laminated sandy silty coarse basal part (TB) and an overlaying normally graded mud part (homogenite, H). The TB part has a sharp and often erosional basal boundary. The sandy silt laminations in the TBs commonly show upward decreasing thickness and grainsize, suggesting deposition from a waning velocity pulse of a turbidity current (Ho et al., 2018; Gutierrez-Pastor et al., 2013). Multi-pulse turbidites are rare or absent in the SoM. Some coseismic THUs consist of stacking of two turbidites, which are likely formed by two different seismic events (e.g. THUs related to the May and Agust 1766 earthquakes), separated by a short time interval, rather than sourced from different canyons.

X-ray radiography, grainsize profiles, MSCL physical properties and μ-XRF elemental profiles provide important insights into the internal structure and geochemical composition of the THUs. The TB parts have high density, MS and commonly high Ca and Sr and low but variable lithofile element (Fe, K, Ti and Zr) contents. The high Ca and Sr are due to the shell debris derived from shallow basinal depths. The variability of the elemental contents are controlled by the heavy mineral and biogenic carbonate composition in the sand-silt fraction of the THUs, which shows changes between different basins. The clayey silty H parts show a gradational upward decrease in grainsize and density, and in elemental profiles, which often show a marked change at their upper boundary with the hemipelagic (HP)sediments.

The coesimic turbidite (THU)-based average earthquake reoccurnce times for the MMF vary from from 220 to 310 years, which is in agreement with historical records, geodetic measurements (2.4 cm/yr; McClusky et al., 2000; Meade et al., 2002), and the lateral offset of the recent earthquake ruptures measured onland ( e.g 4.5–5 m for the 1999 İzmit earthquake by Barka et al., 2002; 1912 Şarköy-Mürefte earthquake by Rockwell, 2001). However, despite this consistency of average recurrence times, the time interval between the consecutive THUs is highly variable, ranging commonly between ~50 and 930 yrs (Fig. 14). The temporal variability in occurrence frequency is likely controlled by an interplay of various factors, including the earthquake magnitude and seismc behaviour the fault segment, core location in relation to basin morphology and earthquake epicentre, SR, sediment sensitivity to earthquake shaking, and paleoceanographic conditions (Yakupoğlu et al., 2022, and references therein). In particular, short average earthquake recurrence interval of 220 years (high occurrence frequency) for the Western Marmara segment is in disagreement with its present creeping (aseismic) behaviour (Ergintav et al., 2014; Schmittbuhl et al., 2016; Bohnhoff et al., 2017; Yamamoto et al., 2018), which likely suggests that the fault may have been episodically locked, and produced more frequent Mw>7 earthquakes in the past periods.

Concerning the effect the core location on the THU frequency, the basin depocentres are the best sites for  deposition of thick and well-defined, but less frequent THUs. This is because these THUs are deposited from relatively large gravity-driven turbidity currents triggered by strong M >7 earthquake, which reach to the depocentre of deep basins.  On the other hand, sites near the fault scarps on the basin margins and sagponds on the highs are sensitive to far-field and local shaking events, and deposit thin and more frequent THUs.

The THUs associated with M>7 earthquakes of  May and August 1766,  1509, 1343-1344, 740 and 557 earthquakes are represented in the different basins and highs in the SoM,  and clearly show their connection with the specific ruptured segment of the MMF (Fig. 14). The May and August 1766 earthquakes shakings were effective in all areas from Çınarcık in the east to the Tekirdağ Basin in the west. The 1509 and 740  earthquakes shakings produced THUs in basins from İzmit Gulf in the east to the Central Basin in the west, but their  records are missing in the Western High and Tekirdağ basin. The distribution of theTHU records of the 1343 and 1344 earthquakes are similar to those of the 1766 earthquakes, and are found all the way from the Çınarcık Basin to the Tekirdağ Basin. The 557 earthquake affected the Çınarcık Basin, Central High, Central Basin. These THU distribution in the SoM strongly suggests that the coesismic turbidites are confined to the basins near the epicentres, and are specific to the ruptured segment.

The last earthquakes with M> 7 recorded as coseismic THU on the different segments in the SoM from east to west are: 1999 İzmit Gulf, 1894 Prince's Islands, 1766 earthquakes in the Kumburgaz and Central basins, and 1923 and 1766 earthquakes on the Western Marmara segment (Western High and Tekirdağ Basin) (Fig. 14). Our coseismic THU records from the Western High and Central Basin indicate that it is unlikely that the rupture of the 1912 Şarköy-Mürefte earthquake extended all the way to the Central Basin. The fact that the record of 1766 (most likely May) earthquake record is the last THU record found on the Central High (Uçarkuş et al., in prep) confirms the locked nature of the Central High segment (Pondard et al., 2007; Ergintav et al., 2014; Schmittbuhl et al., 2016; Bohnhoff et al., 2017), and makes it the most dangerous segement with a potential to create Mw7.2 to 7.4 earthquake in the near future (Parsons, 2004, 2008; Murru et al., 2016).

Another significant result of this study is that the epicentre locations and intensity-magnitude of the earthquakes based on the historical accounts may not always be reliable. This is especially the case for the ancient Greek and early Roman periods, as the records are usually scanty and based on the extent of damage in the few main historical cities (e.g of Lysimachia: Mürefte-Şarköy; Perinthus: Marmara Ereğlisi; Nicomedia: İzmit) that may have been away from epicentres of large earthquakes, but yet suffered some damage.

## Acknowledgements

The cores, were obtained during 2007, 2009 and 2010 cruises, on board R/Vs L'Atalante, Le Suroit, and Urania, which were funded by Marnaut and EC FP7 Esonet projects. We would like to thank the captains, and crews of the vessels and scientific team of the missions, as well as the funding organziations. Thanks are also due to Dursun Acar of ITU-EMCOL for handling of cores and carrying out the MSCL and XRF core scanner analyses. We acknowledge funding by the Turkish

# References (to be checked)

# FIGURE CAPTIONS

Fig. 1. (A) Morphotectonic map of the Sea of Marmara (SoM), showing the Main Marmara Fault (MMF) and locations of cores analysed in this study (stars). The MMF segments and other fault traces are after Le Pichon et al. (2001), Armijo et al. (2002) and Uçarkuş et al (2010). The MMF segments are: IZ= İzmit Gulf, PI=Prince Islands, CH=Central High, WM= Western Marmara, Ga= Ganos. The boxes B-G correspond to the maps of SoM morphotectonic regions along the MMF, shown with the same labels in B-G. The inset map shows the active tectonics of Turkey and surrounding region, with the North Anatolian Fault (NAF), East Anatolian Fault (EAS), European (Eu), Anatolian (An), and Arabian (Ar) plates. Triangles show the location of cores previously studied by Sarı and Çağatay (2006) (blue); McHugh et al. (2006) (green); Çağatay et al. (2012) (pink) McHugh et al. (2014) (orange); Drab et al. (2012) (magenta); Drab et al. (2015) (white); Yakupoğlu et al. (2019, 2022) (yellow).

Fig. 2. Chirp seismic profiles with the studied core locations. (A) C-7 from eastern part of Çınarcık Basin showing the sedimentary units dipping and thickening towards the active Prince Islands Fault in the north. The units thin and wedge out and cut by en-echelon extensional faults towards the southern margin. (B) Profile XYK-02 in the Kumburgaz Basin on Central High, showing flat laying sediments and the projected core locations on the profile. (C) Profile CB-03 across the Central Basin; note the 40 m-high fault scarps on southern margin the rhombic inner basin. (D) Profile T-30 on the Western High, showing location of interface core MEI-02 in a sagpond. (E) Profile T-17 in Tekirdağ Basin showing the thickening of sedimentary units towards the active Western Marmara Fault segment in the southern margin of the basin. See Fig. 1A for location.

Fig. 3. (A) Lithology, X-ray radiography, MSCL physical properties, μ-XRF elemental composition (in z-standardized units) and grainsize-composition profiles of core KI-08 in Çınarcık Basin. Note the presence of two turbidite units (THU), with coarse basal TB (in red) and homogenite (H; in yellow) parts. (B) Lithology and MSCL physical properties profiles of core KI-07 with three THUs. (C) and $^{137}$Cs profile of core MNTKI-7 in Çınarcık Basin, showing the 1986 Chernobyl and 1965 atmospheric nuclear explosions peaks. See Fig. 1C for core locations.

Fig. 4. Lithological description, with calibrated radiocarbon ages, MSCL physical properties, X-ray radiography and μ-XRF elemental composition (in z-standardized units) of core KS-10

in Çınarcık Basin. Coseismic turbidites (THUs) are shown in red (coarse basal part, TB ) and yellow (homogenite, H) in the stratigraphic column. See Fig. 1C for core location.

Fig. 5. Core photographs are showing the sedimentary structures of coseismic turbidites (THUs) in core KS-10 from Çınarcık Basin. (A) A coseismic turbidite-homogenite unit (THU-2) with laminated basal sand layers (TB in red), overlain by a homogeneous mud (homogenite, H in yellow). Note the terrestrial organic debris in the middle laminae (arrow). (B) Multiple sand layers in the basal part (TB) of THU-11, with a dark coarse lamina at the top containing land-derived organic material (arrow), and ~5 cm-thick homogenite part (H), which is overlain by olive green hemipelagic mud (HP). (C) Liquefaction structures: sand lenses and sand balls at ~154 cmbsf and the deformed basal sand layer THU-5. See Fig. 4 for the lithological log and physical and geochemical properties, and Fig. 1B for core location.

Fig. 6. Lithological description, X-ray radiography, MSCL physical properties, μ-XRF elemental composition (in z-standardized units), and grainsize-composition profiles (A), and [137]Cs (B), and [210]Pb (C) profiles of core MEI-01 in Kumburgaz Basin, Central High segment. The uppermost 1 cm of the core is watery, organic-rich dark grey mud, underlain by a 4 cm-thick brown oxidized zone, laminated in the upper part. Coseismic turbidite (THU) is shown in red (coarse basal part) and yellow (homogenite) in the stratigraphic column. See Fig. 1C for core location.

Fig. 7. Lithological description of piston core MEG-02 in Kumburgaz Basin, Central High segment. Note the marine/lacustrine boundary at 1.98 mbsf. Arrows indicate the 3-5 cm authigenic carbonate nodules. This core contains no turbidite-homogenite (THU) unit. See Fig. 1D for core location.

Fig. 8. Lithological description, X-ray radiography, MSCL physical properties, μ-XRF elemental composition (in z-standardized units) (A), [137]Cs (B) and [210]Pb (C) profiles of core KS-18 in Central Basin, with calibrated radiocarbon ages. Coseismic turbidites (THUs) are shown in red (coarse basal part) and yellow (homogenite part) in the stratigraphic column. The grey unit in the upper 14-25 cm interval is the mixed layer layer, which is supported by the presence of measurable [137]Cs down to 30 cm core depth. See Fig. 1E for core location.

Fig. 9. Lithological description, X-ray radiography, MSCL physical properties, μ-XRF elemental composition (in z-standardized units) and grainsize-composition (A) and radionuclide profiles (B) of core MEI-04 on Western High segment. Coseismic turbidites (THUs) are shown in red (coarse basal part) and yellow (homogenite) in the stratigraphic

column. Grey unit in 5-12 cm interval is disturbed layer, tentatively correlated with M7.4 1912 Şarköy-Mürefte earthquake. See Fig. 1F for core location.

Fig. 10. Lithological description, MSCL physical properties, μ-XRF elemental composition (in z-standardized units) (A) and radionuclide profiles (B) of core KI-12 in Tekirdağ Basin. Coseismic turbidites (THUs) are shown in red (coarse base) and yellow (homogenite) in the stratigraphic column. See Fig. 1G for core location.

Fig. 11. Lithological description, μ-XRF elemental composition (in z-standardized units) and MSCL physical properties profiles of core KS-32 in Tekirdağ Basin, with calibrated radiocarbon ages. Coseismic turbidites (THUs) are shown in red (coarse base) and yellow (homogenite) in the stratigraphic column.See Fig. 1F for core location.

Fig. 12. Correlation of the studied cores from Çınarcık Basin, based on lithology, physical properties, and radiocarbon ages. Coseismic turbidites (THUs) are shown in red (coarse base) and yellow (homogenite) in the stratigraphic column. See Fig. 1C for core locations.

Fig. 13. Lithological description, and MSCL physical properties and relative grainsize compositon profiles of core IZ-115 in Karamürsel (Central) Basin of İzmit Gulf. Coseismic turbidites (THUs) are shown in red (coarse base) and yellow (homogenite) in the stratigraphic column (modified from Çağatay et al., 2012) . See Fig. 1B for core location. Note that the THU-1 triggered by the 1999 İzmit earthquake is not fully represented in this core due to artefact in piston core recovery. THU-1 is described from an interface core with undisturbed top from the same location.

Fig. 14. Map showing the distribution of the coseismic turbidites (THUs) correlated with historical earthquakes along the Main Marmara Fault (MMF). Note that the data from this study are integrated with those from McHugh et al., (2006); Çağatay et al., (2012); Drab et al., (2012); McHugh et al., (2014) Yakupoğlu et al. 2019); Uçarkuş et al., (in prep.).

**LIST OF TABLES**

# SUPPLEMENTARY MATRIAL

**List of supplementary figures**

Fig. S1. Non-Bayesian age-depth model for core KS-10 from the Çınarcık Basin, based on AMS radiocarbon datings.

Fig. S2. Non-Bayesian age-depth model for core KS-18 from the Central Basin, based on AMS radiocarbon datings of McHugh et al. (2014).

Fig. S3. Non-Bayesian age-depth model of core KS-32 from the Tekirdağ Basin, based on AMS radiocarbon datings.

**List of supplementary tables**

Tablo S1. Radiocarbon data of cores KI-08 and KS-10 from Çınarcık Basin.

Table S2. Radiocarbon data of core KS-18 Central Basin (McHugh et al., 2014).

Tablo S3. Radiocarbon data of cores KS-32 from Tekirdağ Basin and MNTKI-12 from Western High.

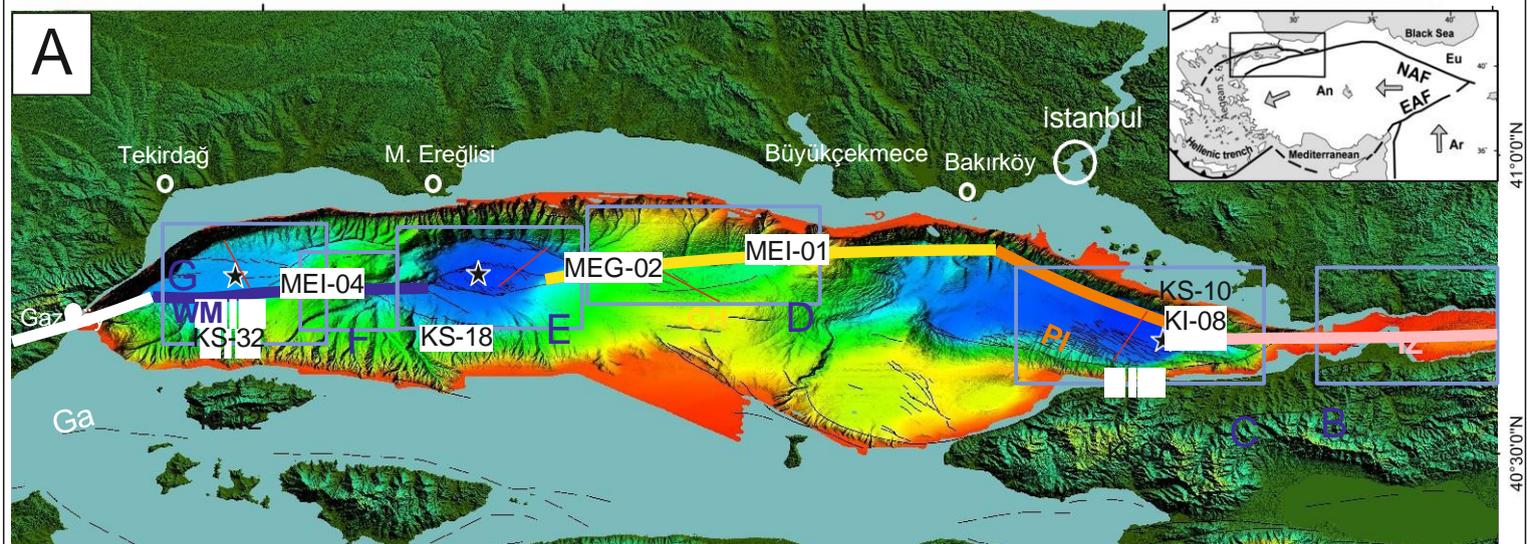

## A

Tekirdağ○  M. Ereğlisi○  Büyükçekmece  Bakırköy  İstanbul

Ga  MEI-04  MEG-02  MEI-01  KS-10
G  KS-32  KS-18  E  D  KI-08
WM  Ga

Topographic elevation (m) 0 — 1200   Bathymetric depth (m) -75 — -500 — -1250

Inset: NAF, EAF, An, Ar, Eu, Black Sea, Mediterranean, Hellenic trench

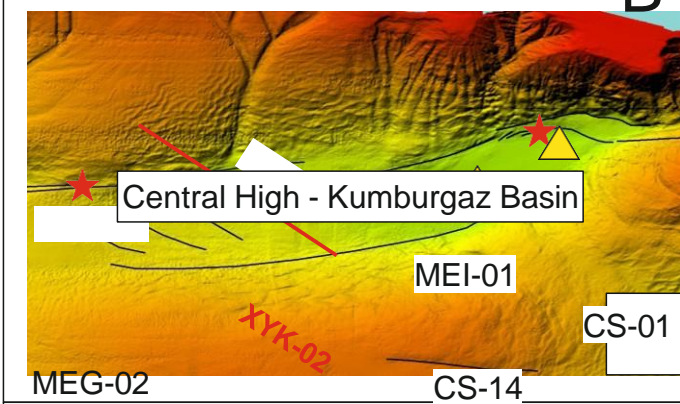

## B

Gulf of İzmit

K1-K2  IZ-113  IZ-115
SW112-113

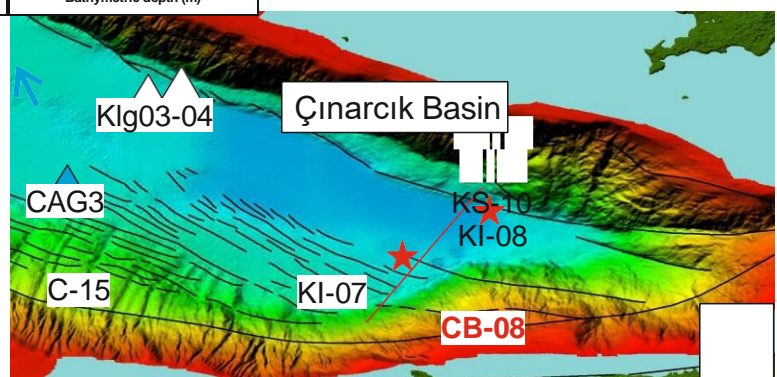

## C

Çınarcık Basin
Klg03-04
CAG3  KS-10
KI-08
C-15  KI-07
CB-08

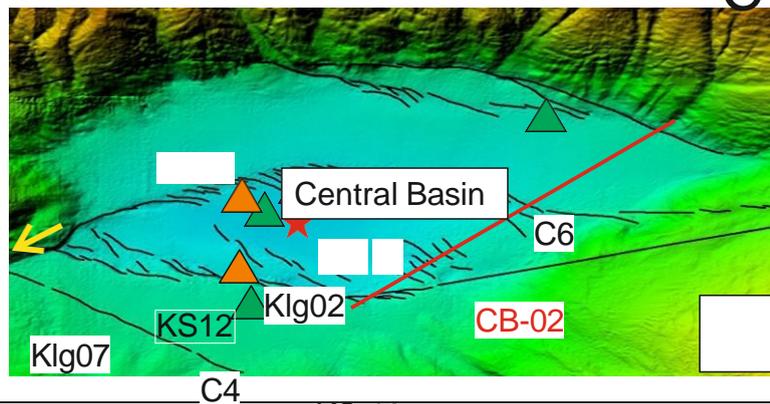

## D / C (Central High)

Central High - Kumburgaz Basin
MEG-02  XYK-02  MEI-01  CS-01
CS-14

Central Basin
Klg07  KS12  Klg02  C6
C4  CB-02

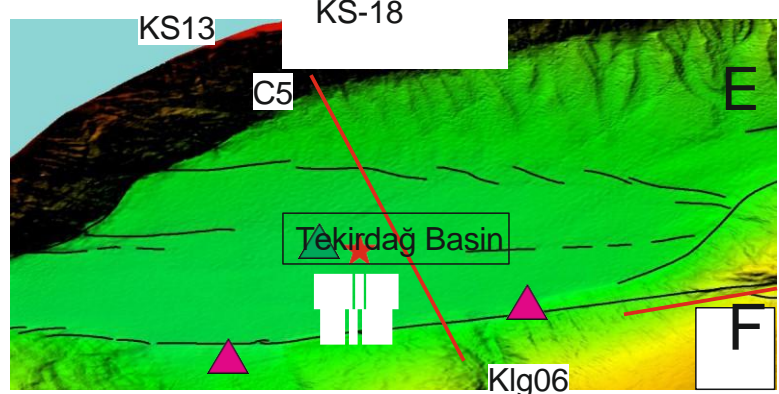

## D / E-F

Western High
MEI-04  T-30

KS13  KS-18  E
C5
Tekirdağ Basin
Klg06  F

C8

KS-32KI-12

Klg05

Klg08

G

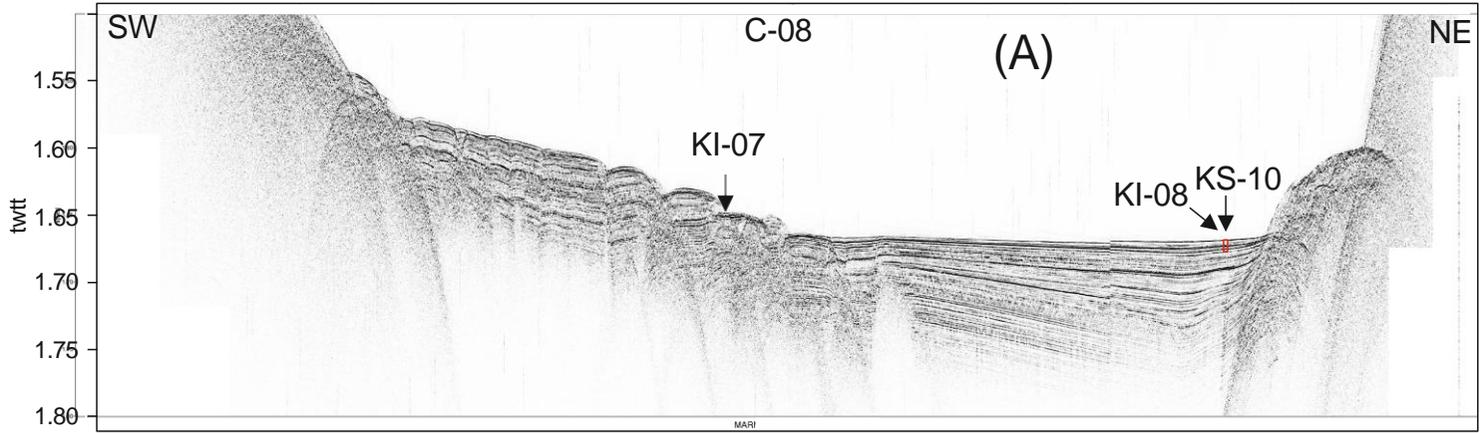

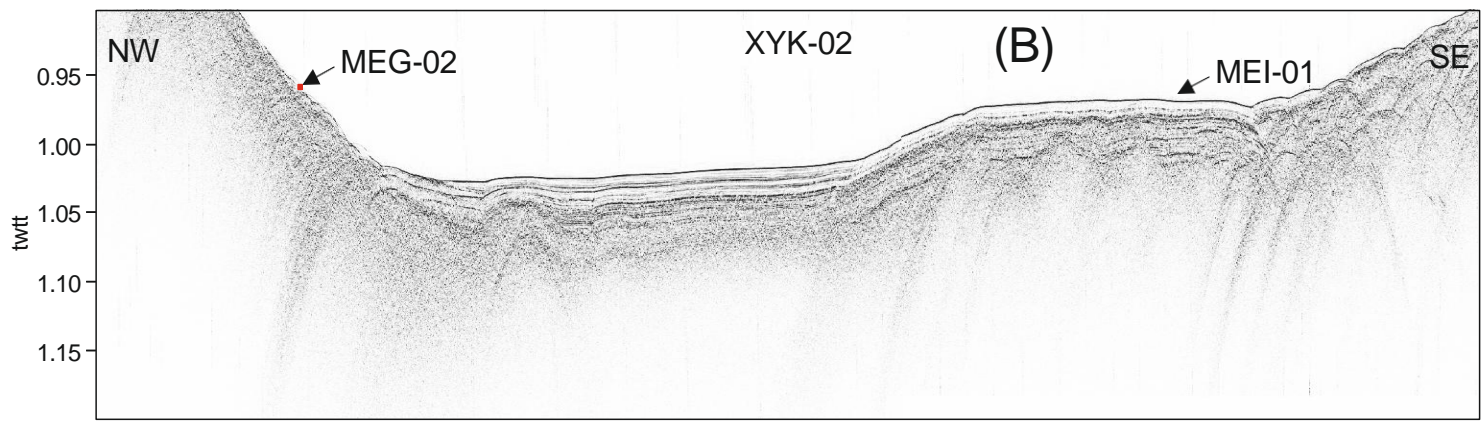

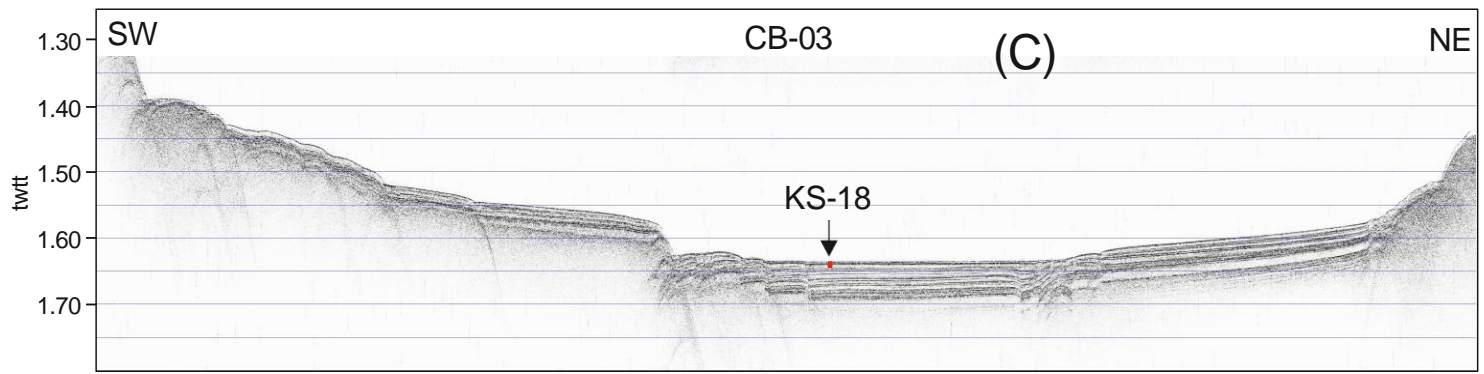

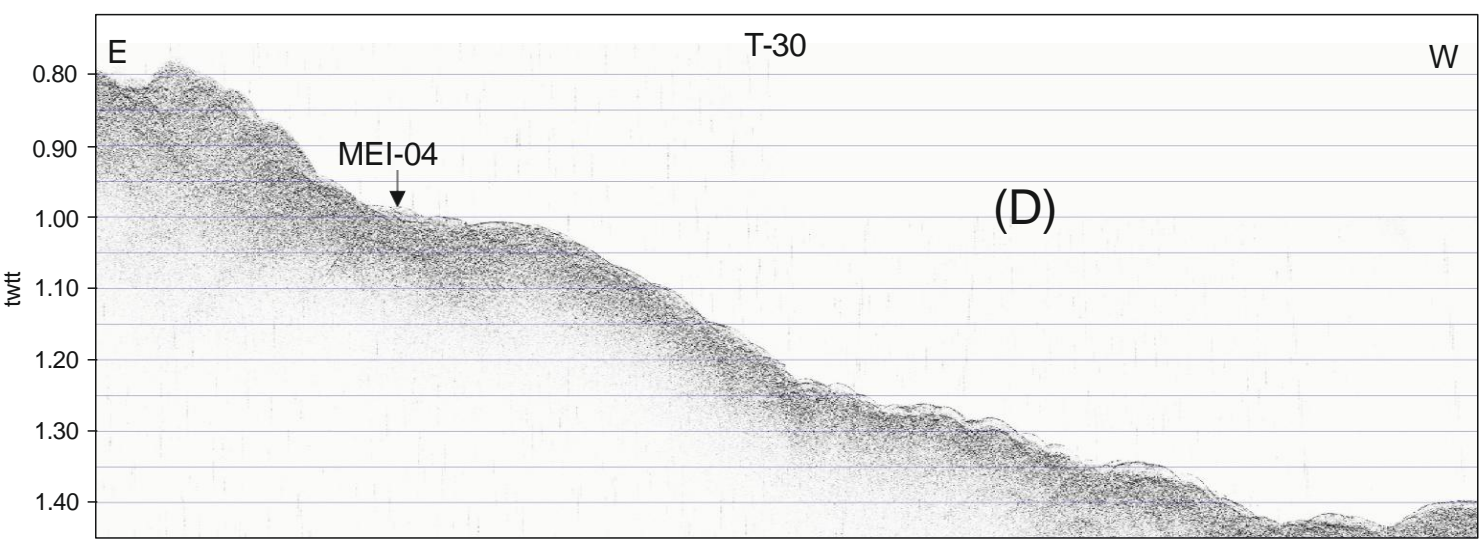

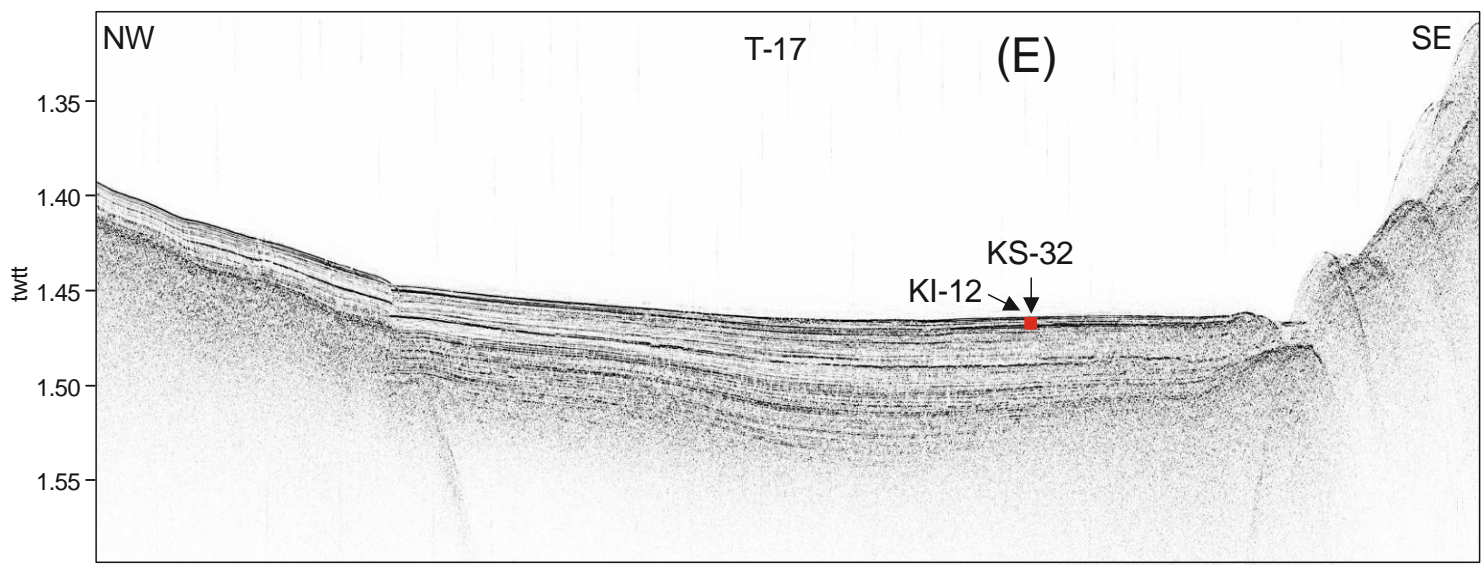

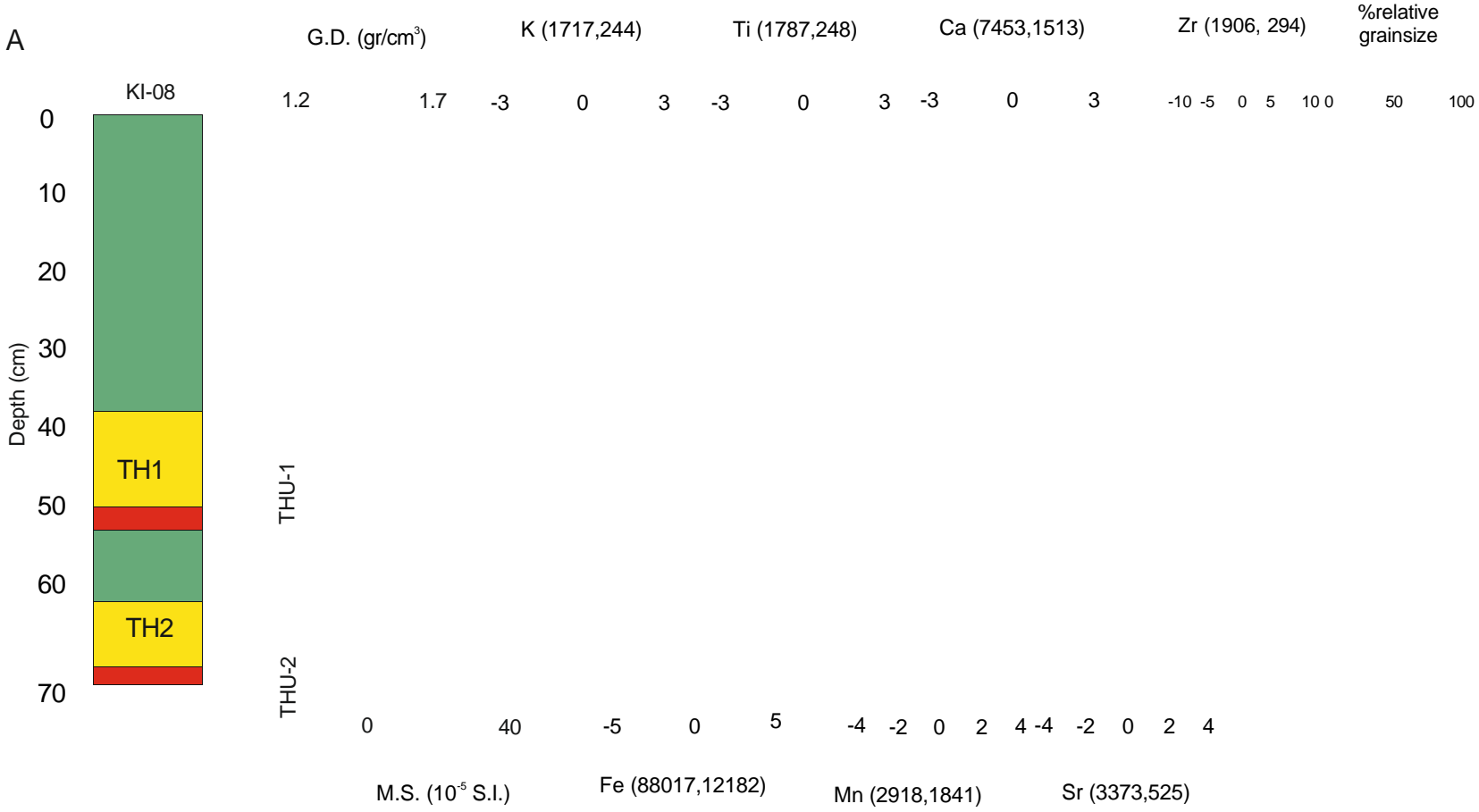

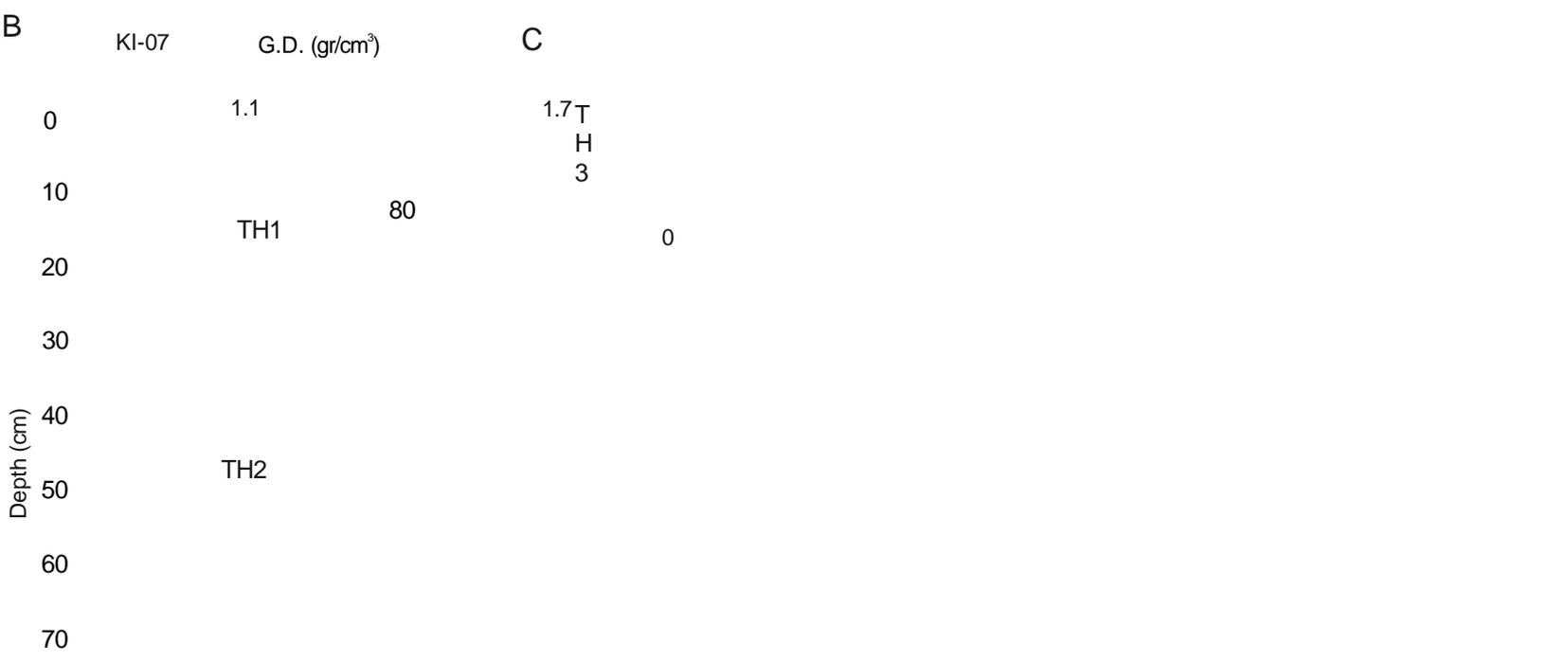

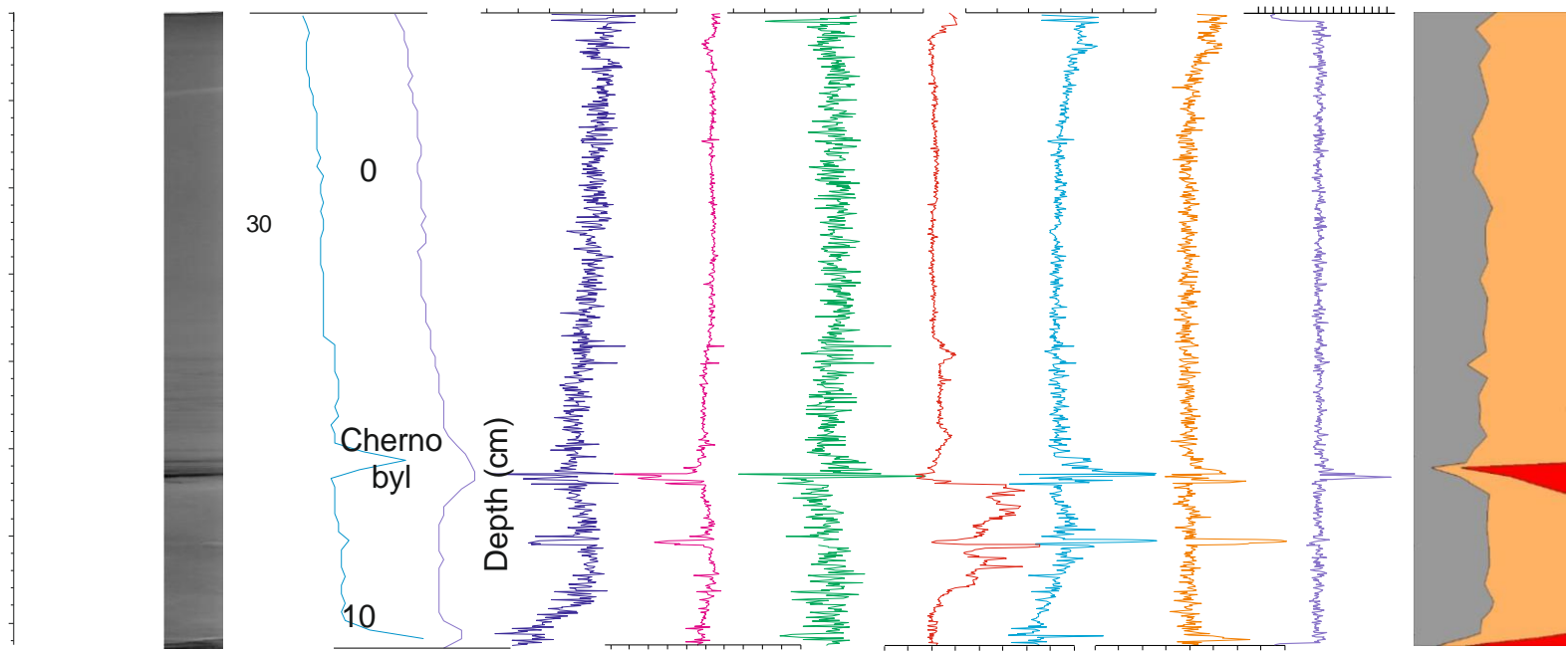

30  0

Cherno byl

Depth (cm)

Nuclear tests

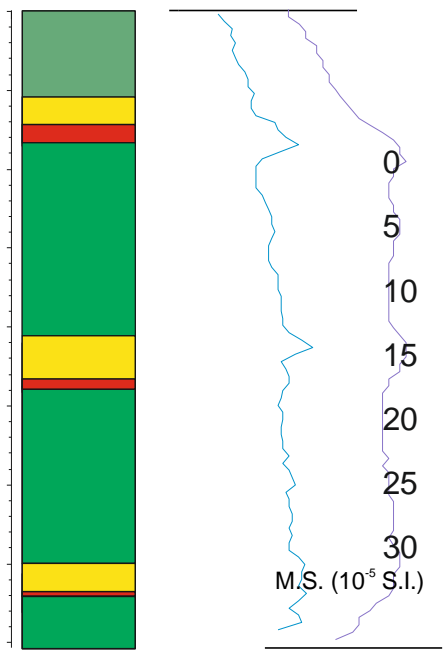

0
5
10
15
20
25
30

M.S. (10$^{-5}$ S.I.)

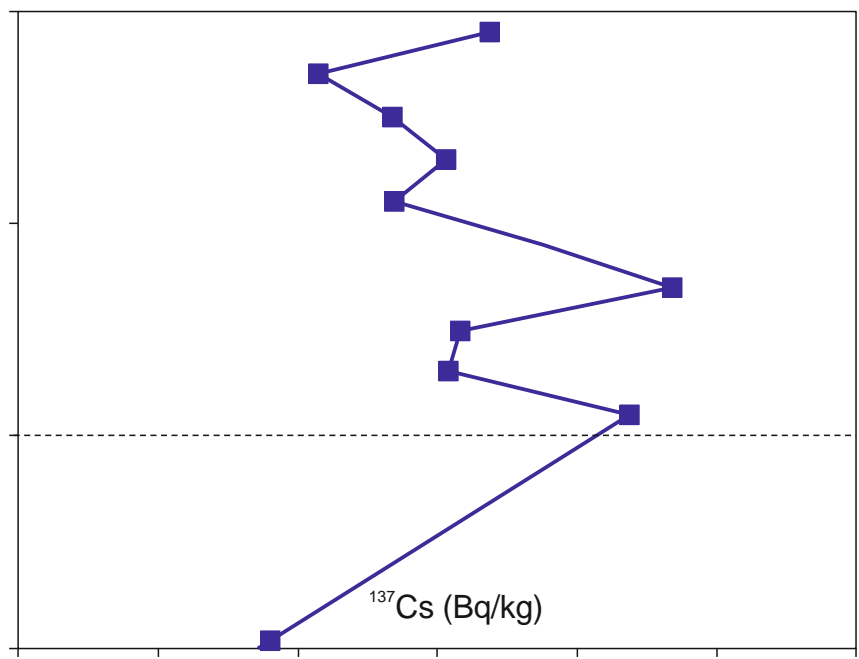

$^{137}$Cs (Bq/kg)

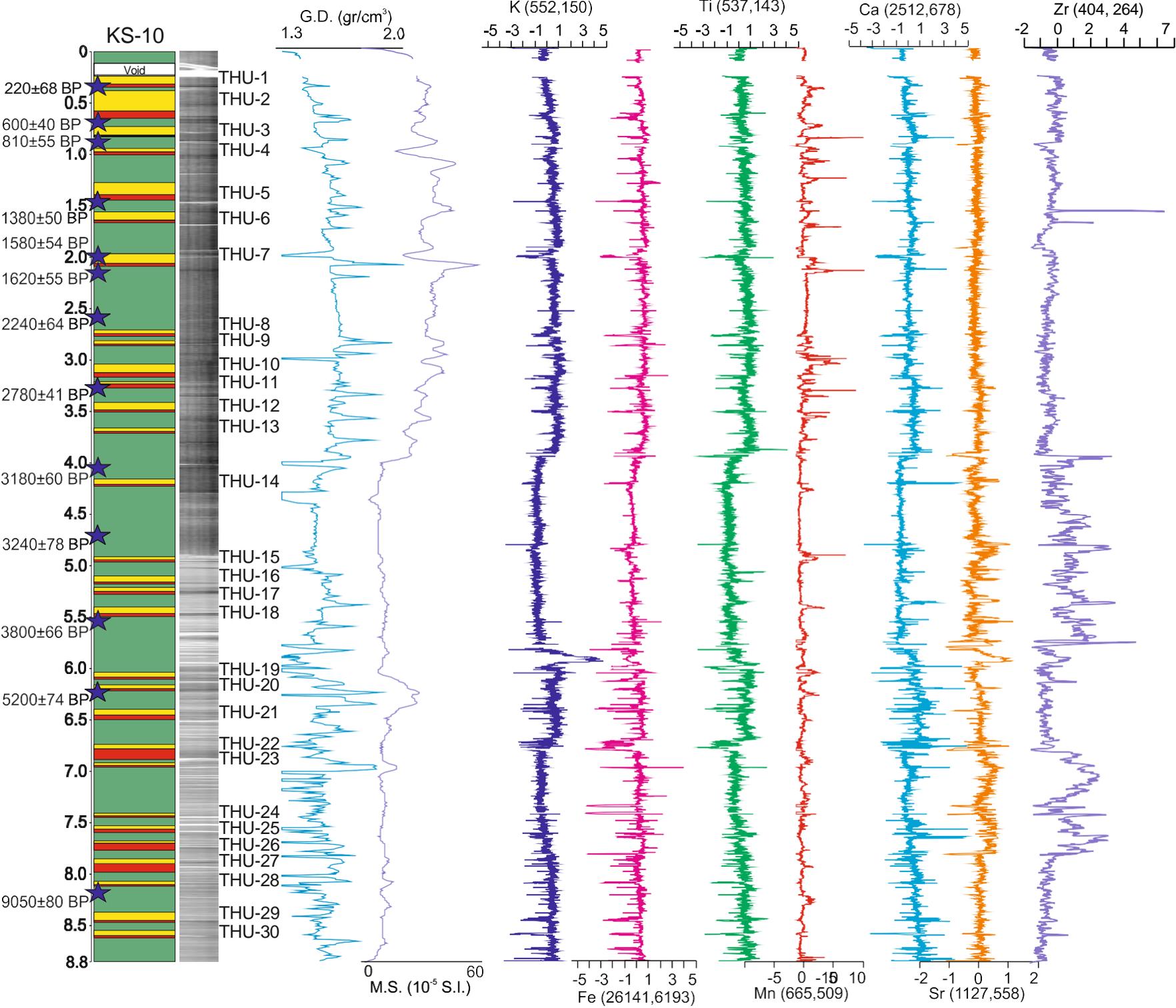

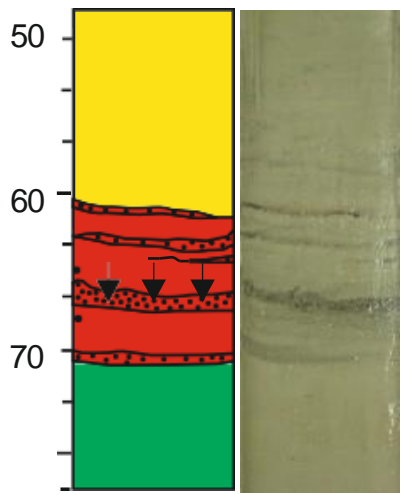

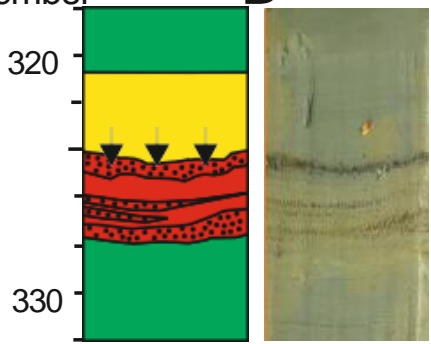

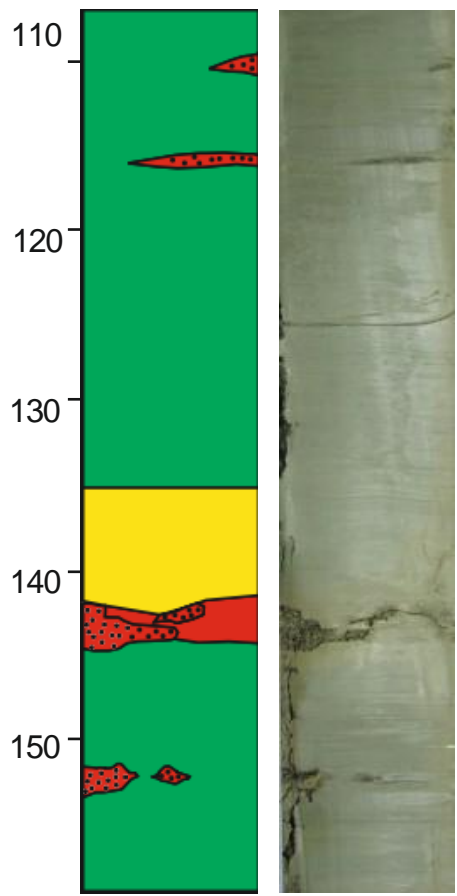

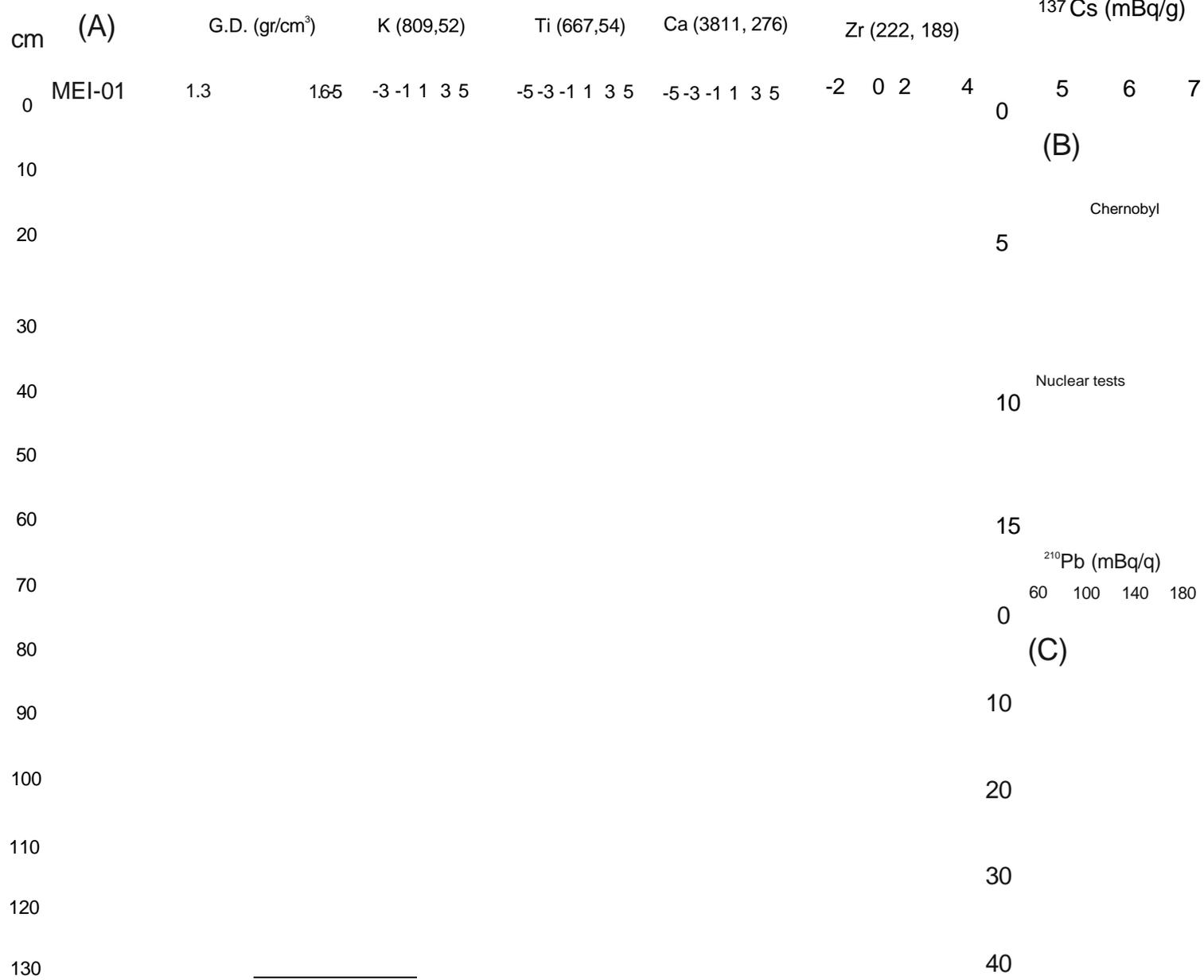

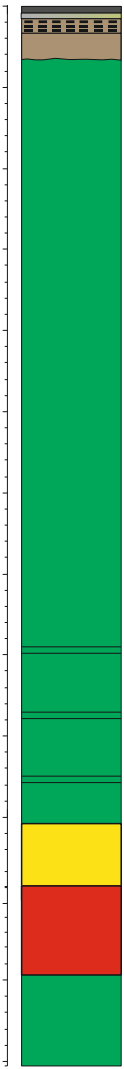
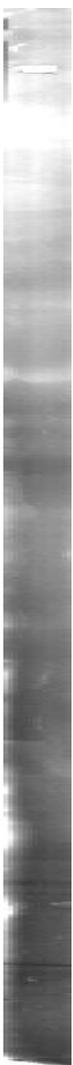
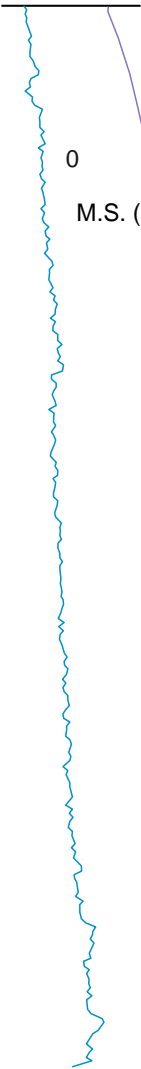
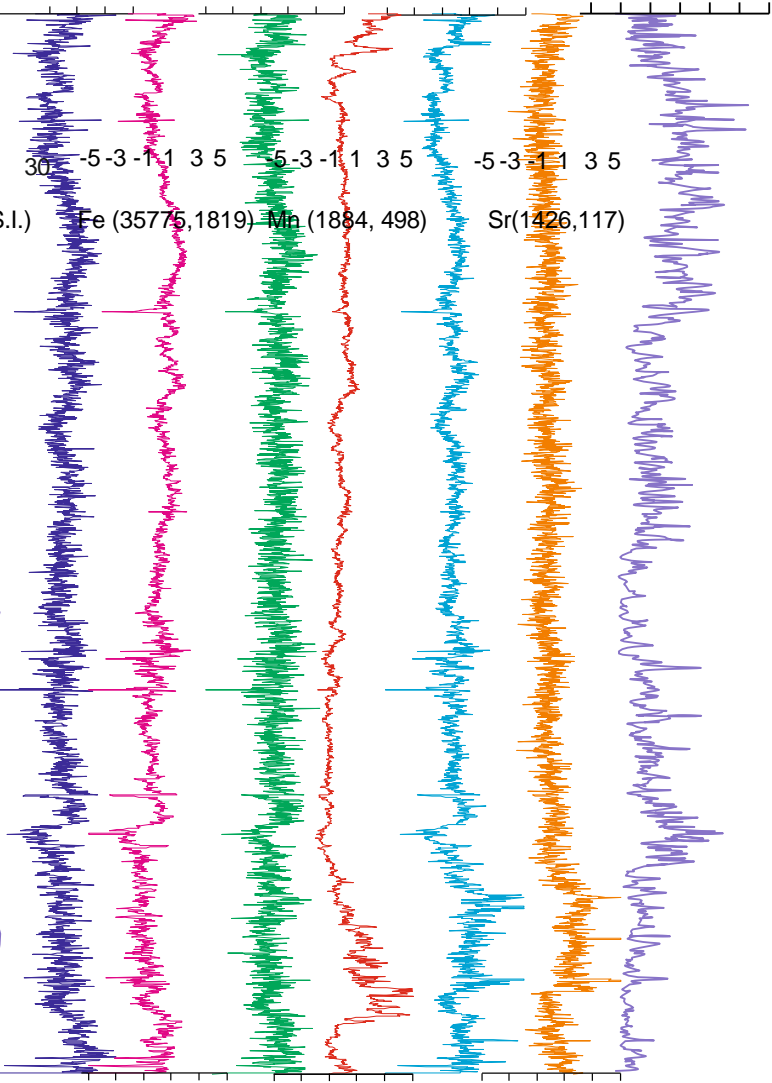
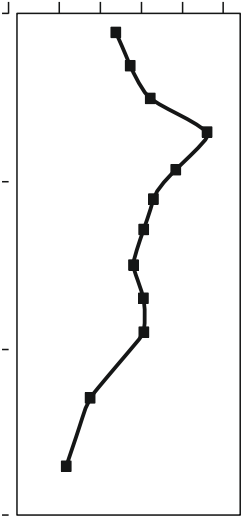
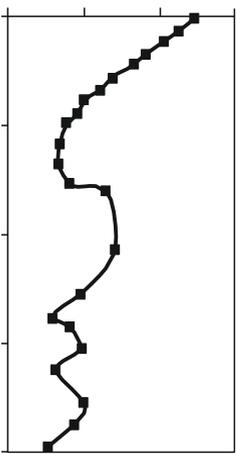

M.S. (10$^{-5}$ S.I.)

0          30

-5 -3 -1 1  3 5    -5 -3 -1 1  3 5    -5 -3 -1 1  3 5

Fe (35775,1819)  Mn (1884, 498)    Sr(1426,117)

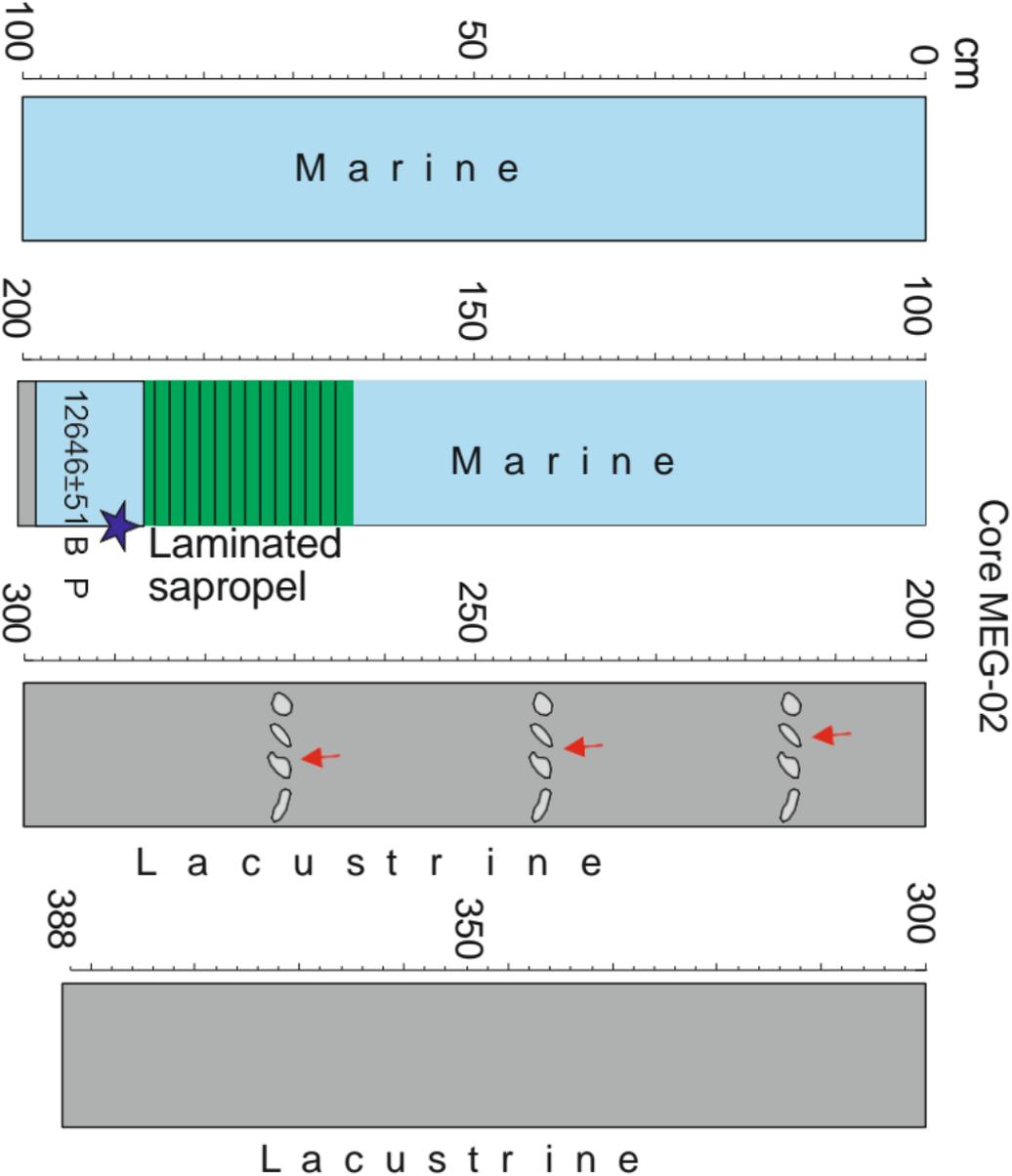

G.D. (gr/cm³)　　　　K (699,61)　　　Ti (572,57)　　　Ca (3065,412)　　Zr (334, 136)

MNTKS-18

1.1　　　1.7

-5 -3 -1 1 3 5　　　-5 -3 -1 1 3 5　　　-5 -3 -1 1 3 5　-5　-3　-1　1 2 3　5

¹³⁷Cs

Depth (cm)

55±50 BP

TH 1
1200±51 BP

TH 2

TH 3

0　　30　　　　　　　　　-5 -3 -1 1 2 3 5　　　　-5 -3 -1 1 1 3 5

TH 4
TH 5
2790±52 BP

TH 6
3435±66 BP
TH 7
TH 8
3620±63 BP
TH 9

3730±66 BP
TH 10

TH 11

TH 12

TH 13
TH 14
4790±85 BP
TH 15
TH 16

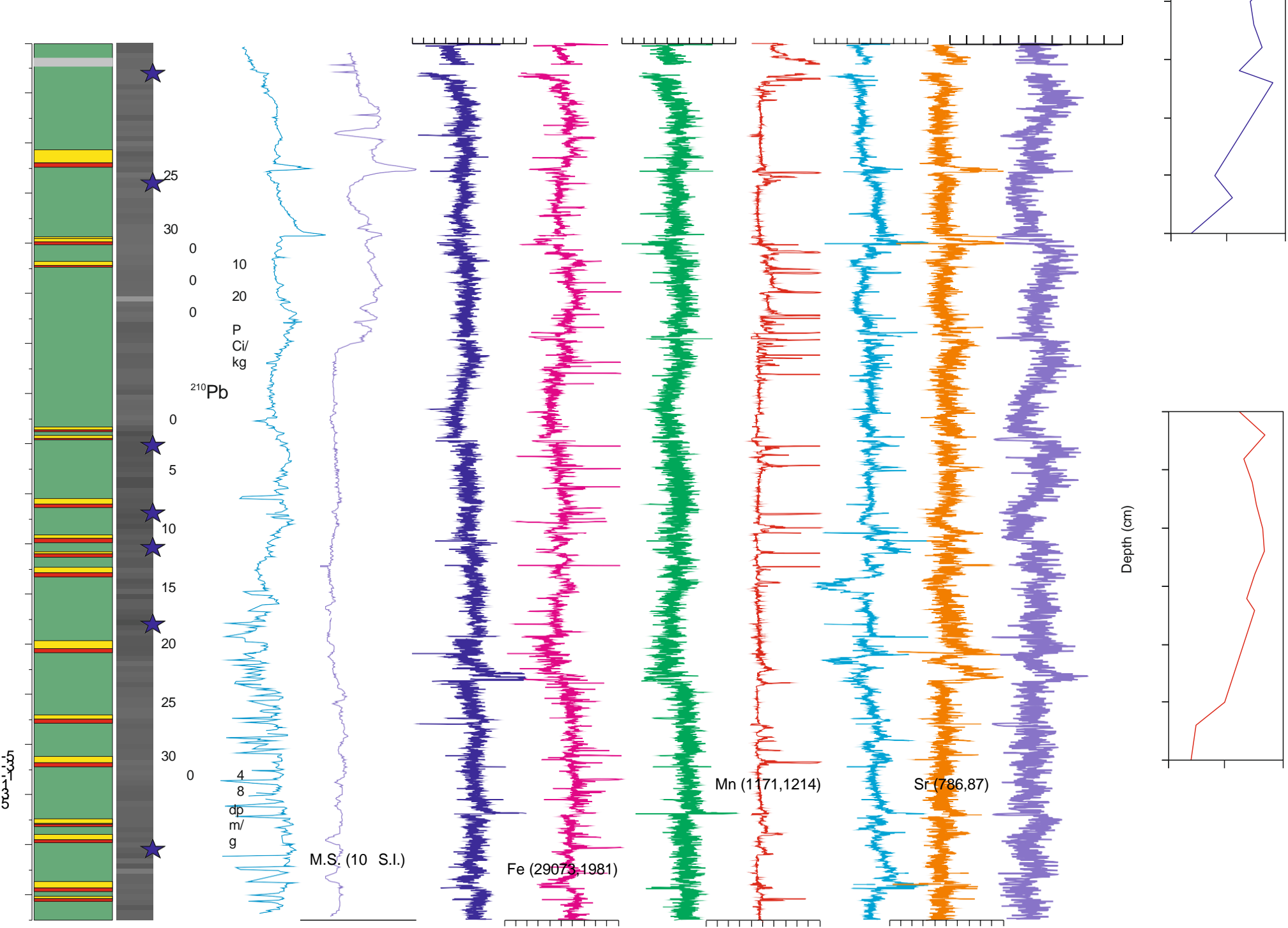

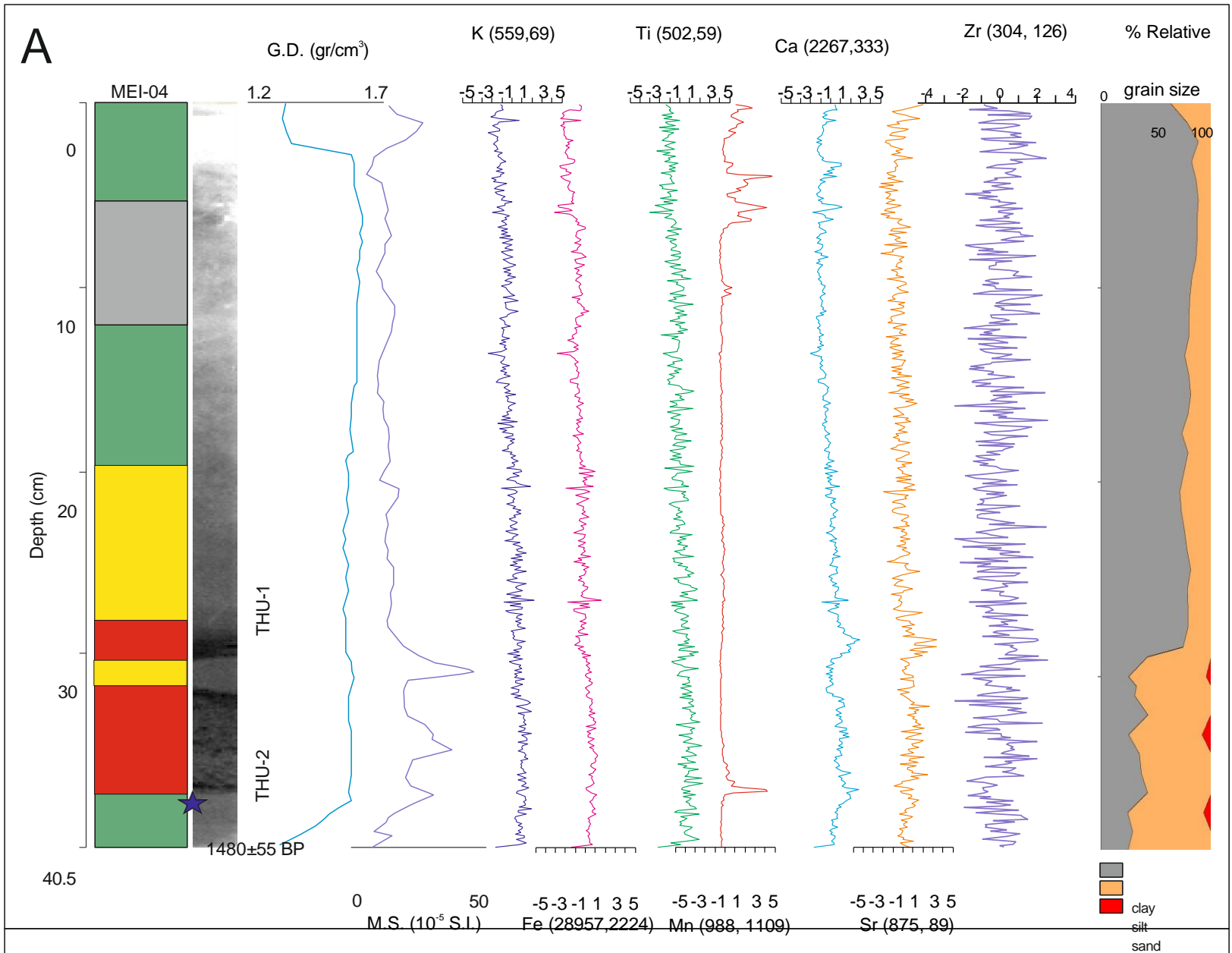

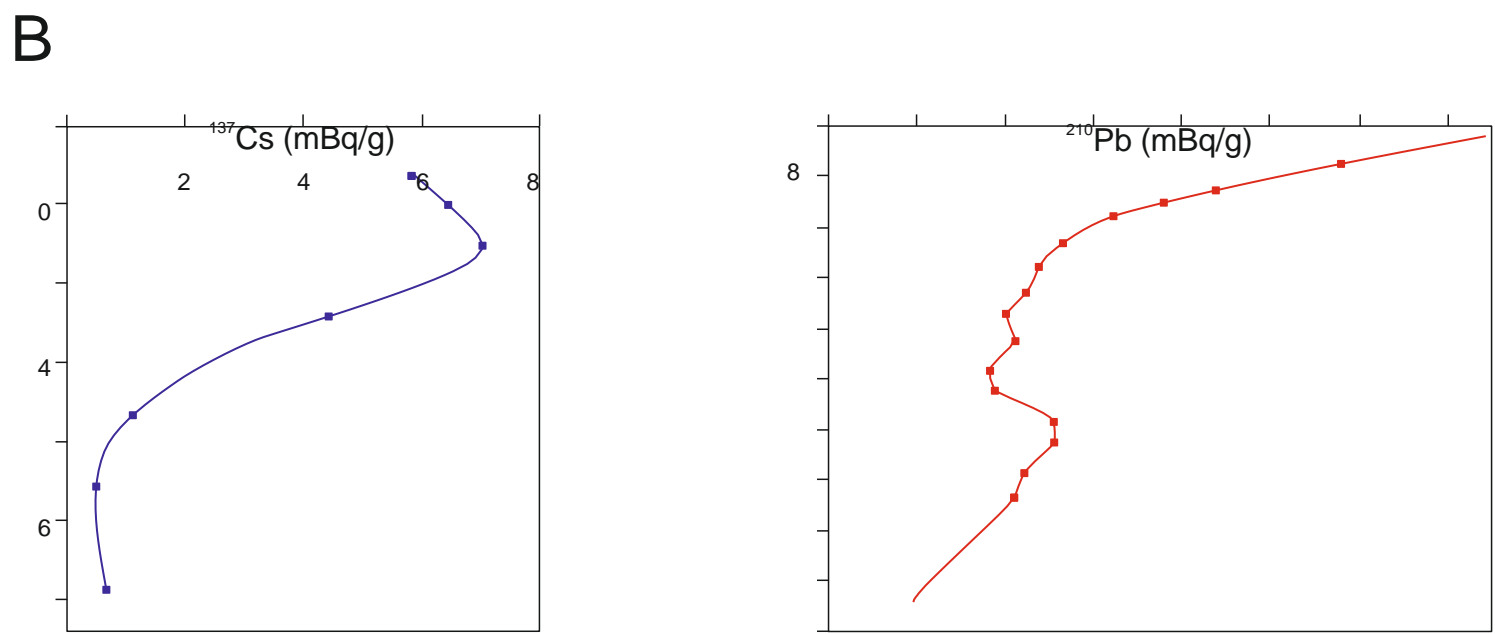

| | 0 | | 40 | | 80 | |
|---|---|---|---|---|---|---|
| | | | 120 | | | |
| 0 | | | | | | |
| 4 | | | | | | |
| 8 | | | | | | |
| 12 | | | | | | |
| 16 | | | | | | |
| 20 | | | | | | |

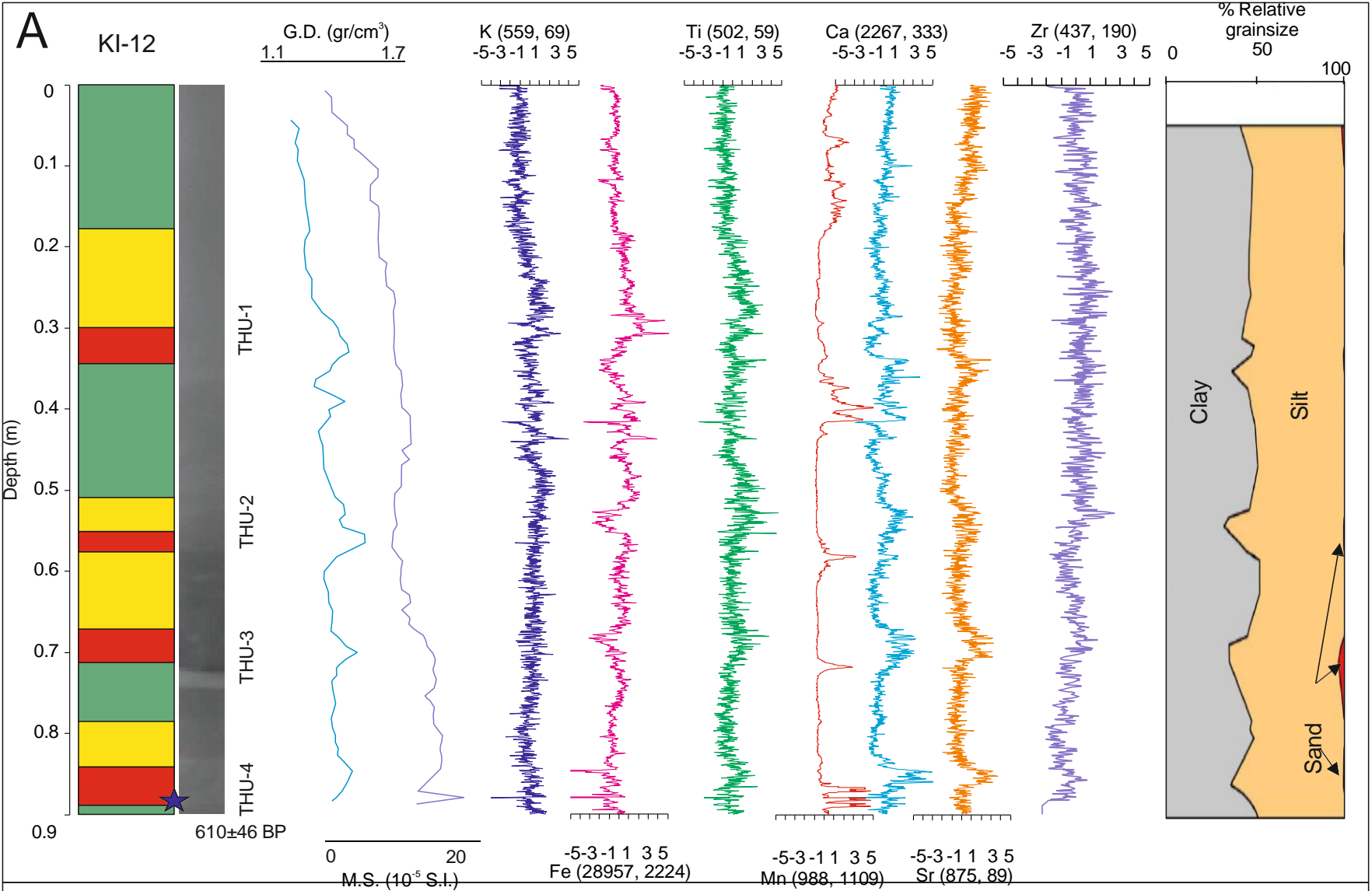

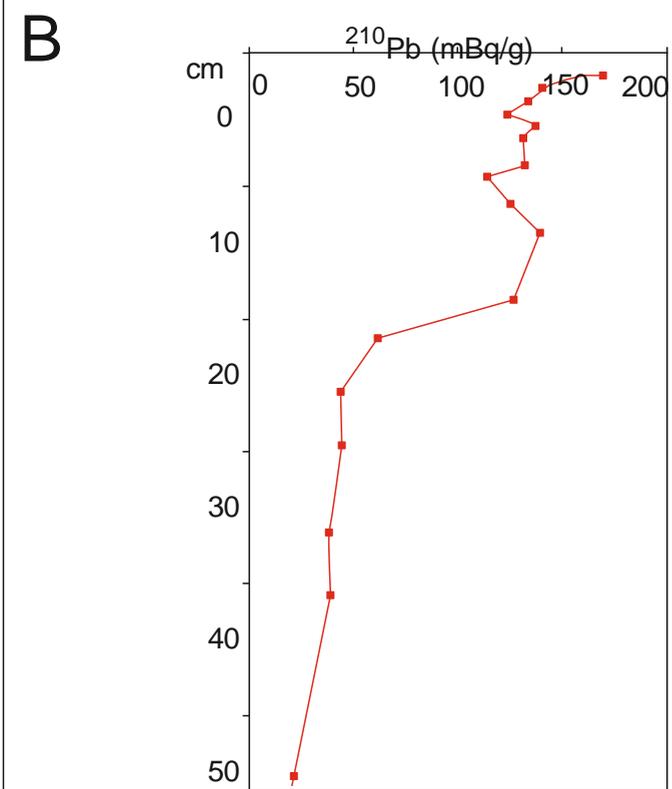

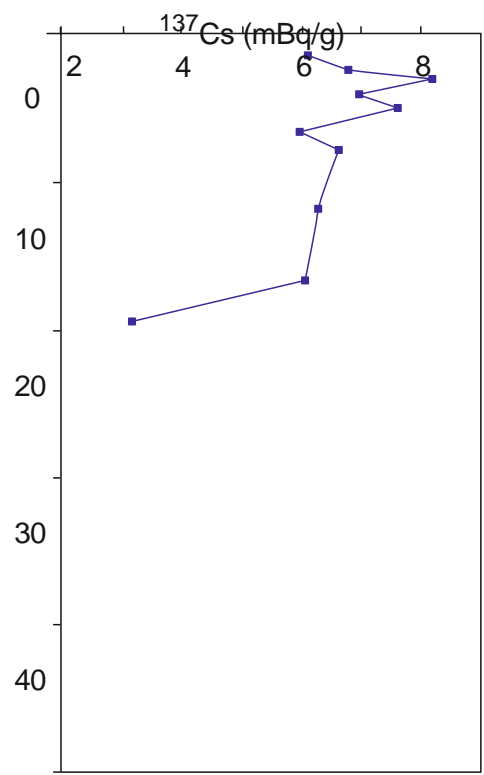

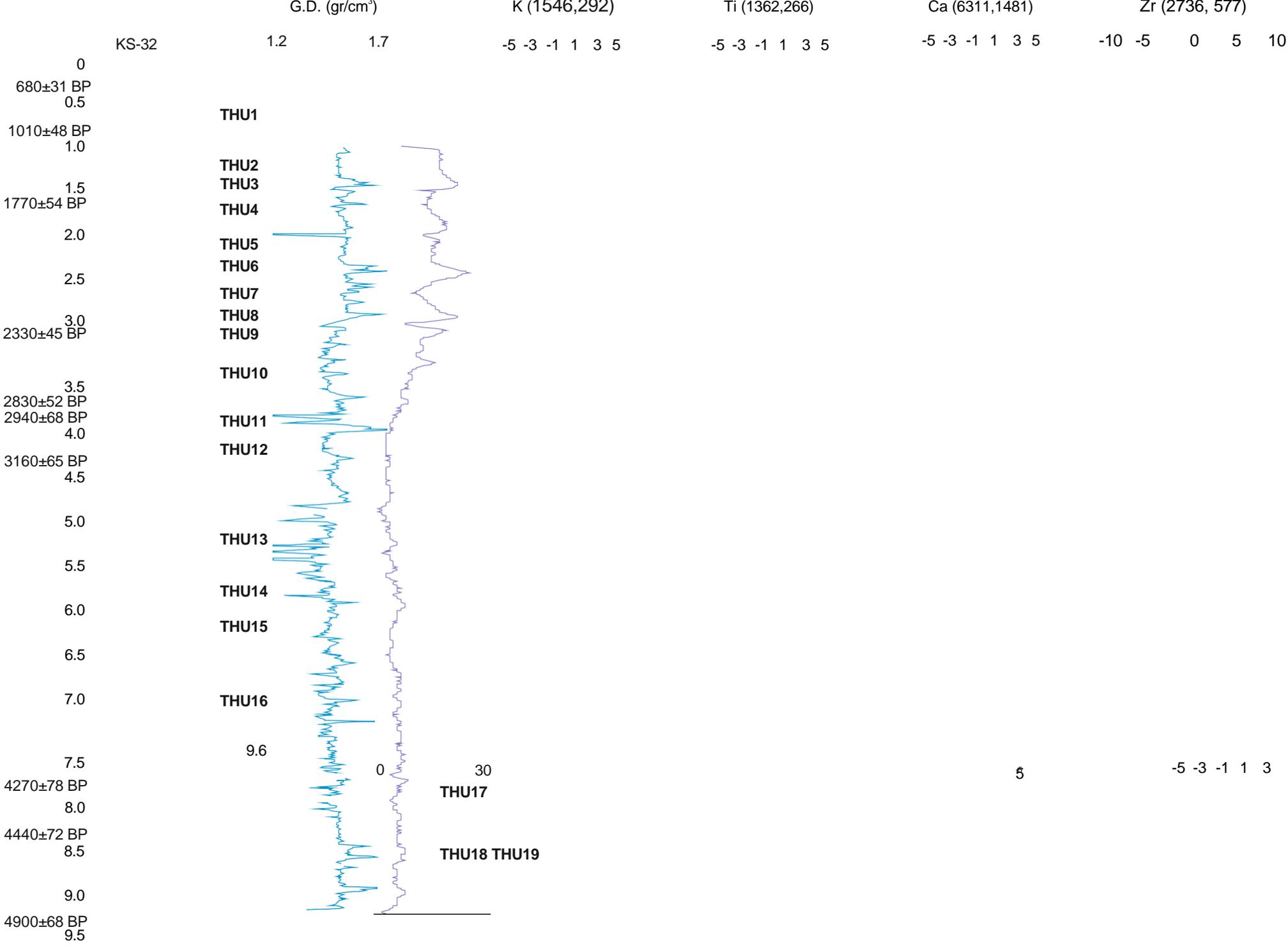

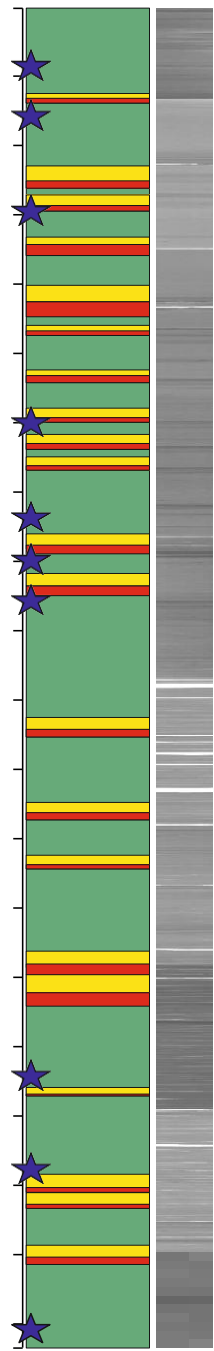
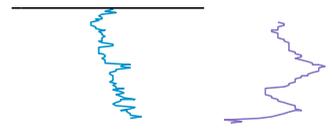
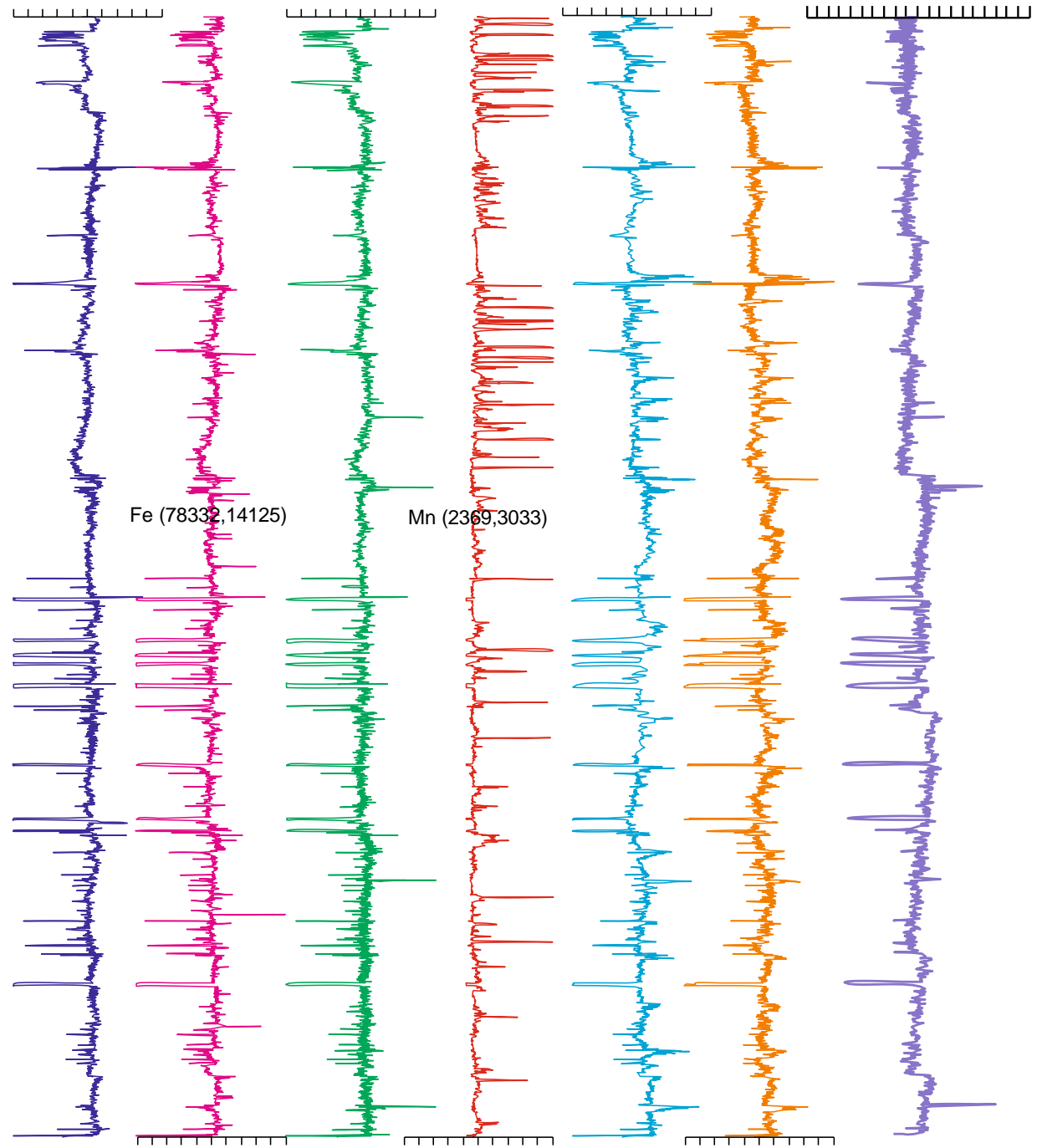

M.S. (10  S.I.)

Fe (78332,14125)

Mn (2369,3033)

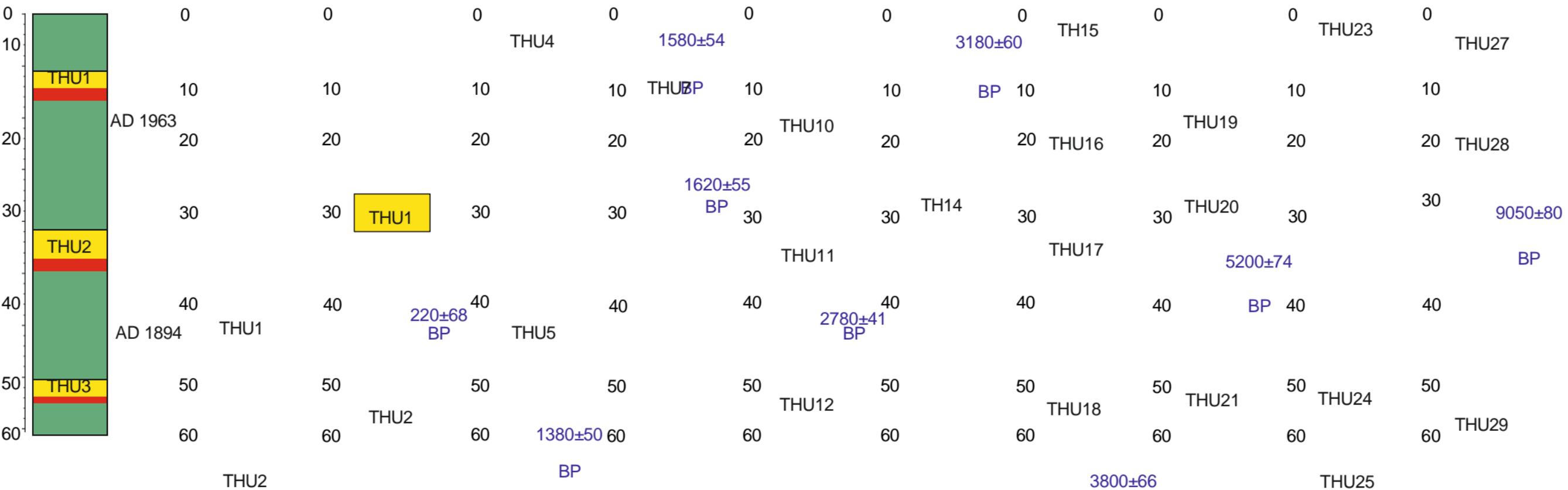

Core I2-115
(-199m waterdepth

Density (g/cc)

Mag. sus. (u.S.I)

Relative%

Core depth

THU-2

THU-3

THU-4

THU-5

Clay

Silt

Sand

Sand

ti:m�:e

1
60

200

220

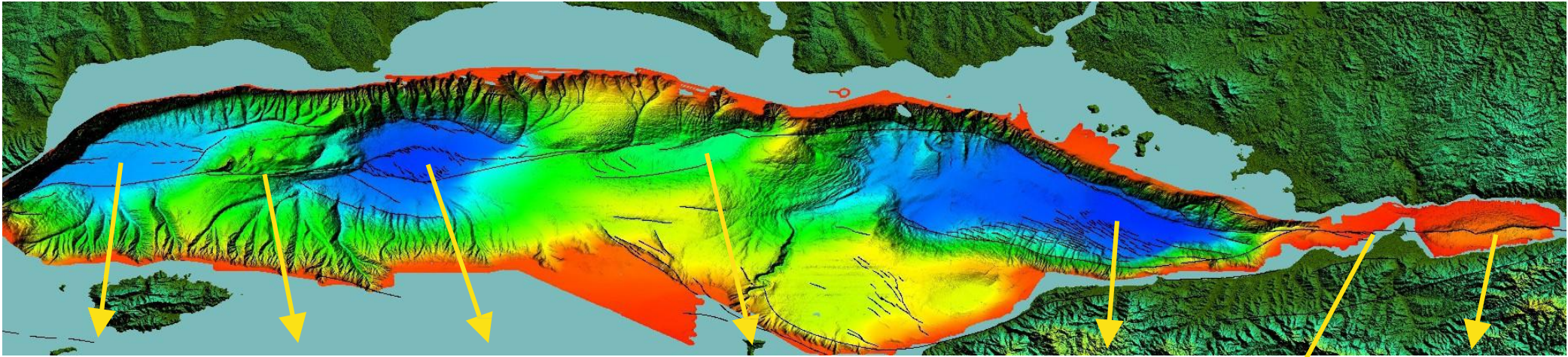

| Tekirdağ Basin |
|---|
| This study |
| AD 1912 |
| AD 1766 |
| AD 1344 |
| AD 989 |
| AD 447 |
| AD 180 |
| AD 120/128 |
| 287 BC |

| Western High |
|---|
| This study |
| AD 1912 |
| AD 1766 |
| Drab et al. (2012) |
| AD 1343 |
| AD 1063 |
| AD 557 |

| Central Basin |
|---|
| McHugh et al. (2014) |
| AD 1766 |
| AD 1509 |
| This study |
| AD 1343 |
| AD 869/945 |
| AD 860 |
| AD 740 |
| AD 557 |
| AD 407 |
| AD 268/343 |

| Central High (Kumburgaz) |
|---|
| This study |
| AD 1343 |
| Uçarkuş et al. (in prep.) |
| AD 1766 |
| AD 1509 |
| Yakupoğlu et al. (2019) |
| AD 989 |
| AD 969/862 |
| AD 740 |
| AD 557 |
| AD 447/478 |
| AD 407 |
| AD 120/180 |
| AD 29/69 |

| Çınarcık Basin |
|---|
| This Study |
| AD 1894 |
| AD 1776 |
| AD 1509 |
| AD 1343 |
| AD 740 |
| AD 557 |
| AD 407 |
| 287 BC |
| 360 BC |

| Western (Darıca) Basin |
|---|
| 287/427 BC |

| Central (Karamürsel) Basin |
|---|
| Çağatay et al. (2012) |
| AD 1999 |
| AD 1509 |
| AD 1296 |
| AD 865 |
| AD 740 |
| AD 358 |
| AD 268 |
| 427 BC |

McHugh et al. (2006)

AD 1894
AD 1766
AD 1509

Table 1. Location, water depth, length and cruise details of the studied piston and sediment/water interface (I) cores.

| Core ID | Latitude | Longitude | Water depth (m) | Core length (m) |
|---------|----------|-----------|-----------------|-----------------|
| KS-10 | 29.1797 | 40.7339 | 1265 | 9.05 |
| KI-08 | 29.1797 | 40.7339 | 1265 | 0.83 |
| KI-07 | 29.11416 | 40.71568 | 1265 | 0.71 |
| MEI-01 | 28.50292 | 40.87069 | 800 | 1.3 |
| MEG-02 | 28.255 | 40.84667 | 692 | 3.86 |
| KS-18 | 28.01385 | 40.8233 | 1260 | 9.2 |
| MEI-04 | 27.72211 | 40.81464 | 741 | 0.4 |
| KS-32 | 27.6113 | 40.82895 | 1123 | 9.45 |
| KI-12 | 27.6113 | 40.82895 | 1123 | 0.94 |

Table 2. Model ages of coseismic turbidite units (THU) units (THU) and correlated historical earthquakes in core KS-10 from Çınarcık Basin, using R-studio and the script "CLAM" (Blaauw, 2010).

| THU | Event free depth (cm) | Min (cal yrs BP) | Max (cal yrs BP) | Mean (cal yrs BP) | Calendar date (AD/BC) | Correlated Historical EQs |
|-----|------|------|------|------|---------|------------|
| 1 | 19 | -8 | 237 | 118 | AD 1889 | AD 1894 |
| 2 | 25 | 157 | 332 | 248 | AD 1759 | AD 1776 |
| 3 | 39 | 389 | 544 | 466 | AD 1541 | AD 1509 |
| 4 | 55 | 530 | 742 | 634 | AD 1373 | AD 1343 |
| 5 | 79 | 1177 | 1368 | 1270 | AD 737 | AD 740 |
| 6 | 90 | 1372 | 1581 | 1477 | AD 530 | AD 557 |
| 7 | 120 | 1441 | 1705 | 1570 | AD 437 | AD 407/358 |
| 8 | 190 | 2184 | 2428 | 2310 | 303 BC | 287 BC |
| 9 | 194 | 2244 | 2480 | 2366 | 359 BC | 360 BC |
| 10 | 211 | 2512 | 2675 | 2593 | 586 BC | 427 BC |
| 11 | 215 | 2566 | 2722 | 2644 | 637 BC | |
| 12 | 231 | 2752 | 2915 | 2834 | 827 BC | |
| 13 | 244 | 2886 | 3047 | 2968 | 961 BC | |
| 14 | 276 | 3076 | 3323 | 3205 | 1198 BC | |
| 15 | 334 | 3266 | 3529 | 3396 | 1389 BC | |
| 16 | 348 | 3317 | 3588 | 3457 | 1450 BC | |
| 17 | 352 | 3341 | 3610 | 3479 | 1472 BC | |
| 18 | 364 | 3421 | 3689 | 3560 | 1553 BC | |
| 19 | 419 | 4254 | 4464 | 4358 | 2351 BC | |
| 20 | 424 | 4360 | 4569 | 4465 | 2458 BC | |
| 21 | 459 | 5144 | 5453 | 5304 | 3297 BC | |
| 22 | 482 | 5691 | 6099 | 5906 | 3899 BC | |
| 23 | 486 | 5787 | 6216 | 6014 | 4007 BC | |
| 24 | 530 | 6959 | 7457 | 7215 | 5208 BC | |
| 25 | 541 | 7269 | 7749 | 7515 | 5508 BC | |
| 26 | 544 | 7354 | 7827 | 7597 | 5590 BC | |
| 27 | 551 | 7546 | 8012 | 7786 | 5779 BC | |
| 28 | 563 | 7888 | 8312 | 8107 | 6100 BC | |
| 29 | 599 | 8864 | 9186 | 9027 | 7020 BC | |
| 30 | 606 | 9010 | 9373 | 9195 | 7188 BC | |

Table 3. Model ages of coseismic turbidite units (THU) and correlated historical earthquakes in core KS-18 from Central Basin, using R-studio and the script "CLAM" (Blaauw, 2010). AMS radiocarbon data used in the modelling are from McHugh et al. (2014).

| THU | Event free depth (cm) | Min (cal yrs BP) | Max (cal yrs BP) | Mean (cal yrs BP) | Mean (cal yrs AD/BC) | Correlated Historical EQs |
|---|---|---|---|---|---|---|
| 1 | 124 | 1221 | 1409 | 1316 | AD 691 | AD 740 |
| 2 | 182 | 1535 | 1869 | 1707 | AD 300 | AD 447 |
| 3 | 198 | 1589 | 1971 | 1784 | AD 233 | AD 358 (?) |
| 4 | 361 | 2771 | 2958 | 2865 | 858 BC | |
| 5 | 364 | 2808 | 2993 | 2898 | 891 BC | |
| 6 | 424 | 3415 | 3609 | 3512 | 1505 BC | |
| 7 | 453 | 3495 | 3803 | 3652 | 1645BC | |
| 8 | 460 | 3504 | 3838 | 3673 | 1666 BC | |
| 9 | 472 | 3522 | 3855 | 3694 | 1687 BC | |
| 10 | 537 | 3625 | 3877 | 3748 | 1741 BC | |
| 11 | 602 | 3821 | 4216 | 4016 | 2009 BC | |
| 12 | 636 | 4029 | 4488 | 4257 | 2250 BC | |
| 13 | 688 | 4623 | 4950 | 4792 | 2785 BC | |
| 14 | 698 | 4680 | 5142 | 4921 | 2914 BC | |
| 15 | 749 | 4634 | 6780 | 5743 | 3736 BC | |
| 16 | 755 | 4610 | 7041 | 5859 | 3852 BC | |

Table 4. Model ages of coseismic turbidite units (THU) and correlated historical earthquakes in core KS-32 from Tekirdağ Basin, using R-studio and the script "CLAM" (Blaauw, 2010).

| THU | Event free depth (cm) | Min (cal yrs BP) | Max (cal yrs BP) | Mean (cal yrs BP) | Mean (cal yrs AD/BC) | Correlated Historical EQs |
|---|---|---|---|---|---|---|
| 1 | 72 | 888 | 1069 | 982 | AD 968 | AD 989 |
| 2 | 109 | 1463 | 1643 | 1550 | AD 400 | AD 447 |
| 3 | 121 | 1619 | 1817 | 1720 | AD 230 | AD 180/268/ 270 |
| 4 | 139 | 1797 | 2040 | 1918 | AD 32 | AD 120-128 |
| 5 | 167 | 1995 | 2236 | 2115 | 165 BC | |
| 6 | 177 | 2050 | 2285 | 2167 | 217 BC | |
| 7 | 199 | 2164 | 2385 | 2276 | 326 BC | |
| 8 | 221 | 2325 | 2503 | 2415 | 465 BC | |
| 9 | 230 | 2412 | 2568 | 2490 | 540 BC | |
| 10 | 237 | 2474 | 2639 | 2554 | 604 BC | |
| 11 | 288 | 2776 | 3000 | 2895 | 945 BC | |
| 12 | 312 | 2965 | 3162 | 3063 | 1113 BC | |
| 13 | 398 | 3235 | 4242 | 3731 | 1781 BC | |
| 14 | 450 | 3455 | 4437 | 3955 | 2005 BC | |
| 15 | 476 | 3594 | 4467 | 4031 | 2081 BC | |
| 16 (a) | 548 | 3907 | 4442 | 4174 | 2224 BC | |
| 17 | 605 | 4118 | 4434 | 4271 | 2321 BC | |
| 18 (a) | 668 | 4304 | 4595 | 4448 | 2498 BC | |
| 19 | 693 | 4415 | 4717 | 4566 | 2616 BC | |